\documentclass[journal]{IEEEtran}
\usepackage{enumerate}
\usepackage{graphicx}
\usepackage{latexsym}
\usepackage{amsmath}
\usepackage{amssymb}
\usepackage{epsfig}
\usepackage{cite}
\usepackage{algorithmic}
\usepackage{overpic}
\usepackage{multirow}
\usepackage[table]{xcolor}
\usepackage{graphicx}
\usepackage{colortbl,array}
\usepackage{multirow,bigdelim}
\usepackage{graphicx}
\usepackage{subfigure}
\usepackage{booktabs}
\usepackage{booktabs,amsmath}
\usepackage{fancyhdr, graphicx, amsmath, amssymb}
\usepackage[linesnumbered,ruled]{algorithm2e}
\usepackage{arydshln}
\usepackage[english]{babel}
\usepackage{xcolor}
\usepackage{booktabs,amsmath}
\usepackage{fancyhdr, graphicx, amsmath, amssymb}
\usepackage{nomencl}

\newtheorem{thm}{Theorem}
\newtheorem{lem}{Lemma}
\newtheorem{rem}{Remark}
\newtheorem{cor}{Corollary}

\newtheorem{defa}{Definition}

\newtheorem{prop}{Proposition}

\begin{document}
%
\title{On Inference of Network Topology and Confirmation Bias in Cyber-Social Networks}
%
%
%

\author{Yanbing~Mao and Emrah~Akyol
\thanks{Y.~Mao and E.~Akyol are with the Department of Electrical and Computer Engineering, Binghamton University--SUNY, Binghamton, NY,
13902 USA (e-mail: ybmao@illinois.edu, eakyol@binghamton.edu). }
\thanks{Parts of the material in this paper were presented in a conference presentation at the 59th Allerton Conference on Communication, Control, and Computing, \cite{mao2019network}.}
}

\maketitle

\begin{abstract}
This paper studies topology inference, from agent states, of a directed cyber-social network with opinion spreading dynamics model that explicitly takes confirmation bias into account. The cyber-social network comprises a set of partially connected directed network of agents at the social level, and a set of information sources at the cyber layer.  The necessary and sufficient conditions for the existence of exact inference solution are characterized.  A method for exact inference, when it is possible, of entire network topology as well as confirmation bias model parameters is proposed for the case where the bias mentioned earlier follows a piece-wise linear model. The particular case of no confirmation bias is analyzed in detail. For the setting where the model of confirmation bias is unknown, an algorithm that approximates the network topology, building  on the exact inference method, is presented. This algorithm can exactly infer the weighted communication from the neighbors to the non-followers of information sources. Numerical simulations demonstrate the effectiveness of the proposed methods for different scenarios.
\end{abstract}


\IEEEpeerreviewmaketitle

\section{Introduction}


The inference of network topology from observed state data in dynamical systems is a key problem in several fields ranging from bioinformatics \cite{d2000genetic} to communication\cite{mao2007wireless} and social networks \cite{al2011survey}, see e.g., \cite{brugere2018network} for a comprehensive overview.  We categorize the prior approaches to this  problem into two groups: approximate and exact inferences.

In several practical scenarios,  data to infer the network topology is only partially available or it is stochastic (e.g., noisy), which  transforms the network inference problem into an instance of well-studied estimation problems, with assumptions on  network dynamics. For example, in \cite{ioannidis2019semi}, this problem is studied in the context of structural equation models, while in \cite{nozari2017network}, autoregressive models are employed. The solution approaches in the literature utilize tools from Bayesian analysis and estimation theory \cite{sardellitti2019graph,materassi2010topological}, adaptive feedback control \cite{yu2006estimating, fazlyab2014robust},
 compressed sensing \cite{hayden2016sparse}, or more generally optimization methods with sparsity constraints\cite{segarra2017network, pasdeloup2017characterization}.

In various other settings,  exact topology inference is possible. For example, in \cite{nabi2012network}, a ``node knockout" method is proposed where selected nodes are grounded (set to zero) to identify the network structure.  In \cite{shahrampour2014topology}, this approach is coupled with power spectral analysis with the knowledge of eigenvalue-eigenvector of matrix that describes network structure. We note that while these methods \cite{nabi2012network,shahrampour2014topology} provide exact topology inference, they require the capability of altering (controlling) every node value in the network, which is difficult in practice, if not impossible, for several realistic scenarios, including the case of social networks.

Perhaps closest to the proposed approach here, in \cite{WHT10}, an exact topology inference strategy is presented, primarily for continuous-time consensus dynamics, by transforming the problem into a solution of Lyapunov equation whose numerical solutions are well studied, see e.g., the \cite{bartels1972solution, haber2016sparse}. This approach, unlike the ones in \cite{sardellitti2019graph,ioannidis2019semi,materassi2010topological, yu2006estimating,fazlyab2014robust, nozari2017network,shahrampour2014topology, nabi2012network}, does not require the capability of external stimulation for every node in the network, hence it is potentially applicable to social networks.  However, as we demonstrate later in this paper, this approach is not sufficient for exact inference in the {\it directed} network topologies that we consider here.

In this paper, we focus on topology inference of  social networks. Mathematical models of  opinion evolution over social networks have gathered significant interest from different disciplines, ranging from computer science, control theory to social science, see e.g., \cite{degroot1974reaching,abelson1964mathematical, friedkin1990social,hegselmann2002opinion,acemoglu2011opinion,easley2010networks}. Notably,  in \cite{degroot1974reaching}, every individual updates her belief as an average of her network neighbors. In \cite{friedkin1990social}, the model further involves innate opinions determined by socio-economic conditions in which that individual lives in. Recently,  several variations have been proposed based on these seminal works to more realistically model capture today's social networks, see e.g. the excellent overviews \cite{proskurnikov2017tutorial,proskurnikov2018tutorial,friedkin2011social} and the reference therein.   In this paper, we focus on the topology inference problem of networks with known information dynamics such as the ones in \cite{degroot1974reaching,friedkin1990social}, with the important explicit consideration of a key cognitive bias, known as the {\it confirmation bias}.

Confirmation bias of an individual refers to favoring information which confirms her previously existing beliefs \cite{nickerson1998confirmation}. This bias plays a key role in creating so called ``echo chambers" in social media, where individuals are exposed to only  their side of the story \cite{lazer2018science}. This is partly due to wide use of machine learning algorithms that filter news on social media newsfeeds, such as the ones in Twitter and Facebook. These algorithms automatically utilize and  foster this bias, i.e., present the user only the news that they would like to see, hence contribute to the polarization of public opinion \cite{waldrop2017news, del2016spreading, wsj,cnn}. Partly due to its role in the spread of misinformation, confirmation bias has recently gained a revived interest \cite{waldrop2017news,lazer2018science,bakir2018fake}. This constitutes our primary motivation to study the specific impact of confirmation bias on several aspects of social networks. We consider this work as a part of a comprehensive exploration of mathematical underpinnings of the misinformation spread and polarization, among other complementary studies from our lab, see e.g., \cite{akyol2017information, allerton17,asilomar18,mao2018spread,asilomar,allerton}.

In \cite{mao2018spread}, opinion dynamics in social networks is studied with a particular focus on confirmation bias. Here, the cyber-social network comprises a social layer (individuals) and a cyber layer (information sources or  ``stubborn individuals" who do not change their opinions). The confirmation bias is modeled as a function of the distance between the opinions of individuals and information sources,  explicitly taken into account in the dynamics model. We note that the well-known Hegselmann-Krause model~\cite{hegselmann2002opinion} and its recent variations  \cite{mirtabatabaei2012opinion} also address this bias, where an individual completely ignores the opinions that are ``too far" from hers. This model seems less amenable to detailed analysis than that in \cite{mao2018spread}, which is adopted in this paper.

In this paper, building on the opinion dynamics model in \cite{mao2018spread} and our preliminary analysis reported in \cite{allerton},
we investigate the problem of network topology inference in conjunction with confirmation bias. To the best of our knowledge, this is the first work in the literature (in social sciences as well as in engineering and computer science) that considers the challenges that confirmation bias brings into the  problem of network topology inference, as well as to the problem of joint inference of network topology and bias parameters from opinion observations. Our contributions are summarized as follows.
\begin{itemize}
  \item We characterize the necessary and sufficient conditions for solvability of exact inference of  network topology and  bias parameters, in the case of piece-wise linear bias model with controlled information sources.
  \item Building on the aforementioned characterization, we provide procedures to obtain
  \begin{itemize}
  \item the exact inference of network topology and  bias parameters when the information sources are controlled;
  \item the exact inference of network topology when the information sources are uncontrollable and there is no  bias.
  \end{itemize}
  \item We provide an approximate inference methodology for the case where the information sources are uncontrollable and the bias model is unknown. The proposed method exactly infers partial network topology.
\end{itemize}

This paper is organized as follows. In Section II, we present the notation, a detailed analysis of the relevant prior work, as well as three inference problem formulations. In Sections III, IV and V, we study inference problems I, II and III respectively. We next present our numerical results in Section VI. We finally discuss our conclusions and future research directions in Section VII.

\section{Preliminaries}
\subsection{Notation}
\label{notation}
We let $\mathbb{R}^{n}$ and $\mathbb{R}^{m \times n}$ denote the set of $\emph{n}$-dimensional real vectors and the set of $m \times n$-dimensional real matrices, respectively. $\mathbb{N}$ stands for the set of natural numbers, and $\mathbb{N}_{0} = \mathbb{N} \bigcup {0}$. We define $\mathbf{I}$ and $\mathbf{O}$ as the identity and zero matrices with proper dimension, respectively. Moreover,  we let $\mathbf{1}$ and $\mathbf{0}$ denote the vectors of all ones and all zeros with proper dimension, respectively. The superscript `$\top$' stands for the matrix transposition. For a matrix $\textbf{W} \in \mathbb{R}^{n \times n}$, ${\left[ \textbf{W} \right]_{i,j}}$ and ${\left[ \textbf{W} \right]_{i,:}}$ denote the element in row $i$ and column $j$ and the $i^{\emph{\emph{th}}}$ row, respectively. $\textbf{A} \in \mathbb{R}^{n \times n}$ is a row stochastic  matrix if
$$[\textbf{A}]_{i,j} \ge 0  \,\, \text{and}\,\,  \sum\limits_{j = 1}^n {{{[\textbf{A}]}_{i,j}}}  = 1, ~~i=1,2,\ldots,n.$$ Other important notations are highlighted as follows:
\begin{description}
  \item[$\mathbb{N}_{i}:$]~~~neighbors of individual $\mathrm{v}_{\mathrm{i}}$;
  \item[$\ker(\textbf{S}):$]~~~set $\left\{ {\textbf{y}: \textbf{S}\textbf{y} = {\mathbf{0}}}, \textbf{S} \in \mathbb{R}^{n \times n} \right\}$;
  \item[${\textbf{A}^{ - 1}}\mathbb{O}:$] ~~~set $\left\{ {\textbf{x}: \textbf{A}\textbf{x} \in \mathbb{O}} \right\}$;
  \item[${E_U}(\cdot):$] ~~~expectation over uniform distribution $U(0,1)$;
  \item[$|\cdot|:$] ~~~~(element-wise) modulus of a real (matrix) number;
  \item[$|\mathbb{K}|:$] ~~~cardinality (i.e., size) of a set $\mathbb{K}$.
\end{description}

The network considered in this paper is composed of $n$ individuals (the social part of the network) and $m$ information sources (the cyber part of the network). The interaction among the individuals is modeled by a digraph $\mathfrak{G} = (\mathbb{V}, \mathbb{E})$, where $\mathbb{V}$ = $\left\{\mathrm{v}_{1}, \ldots,  \mathrm{v}_{\mathrm{n}}\right\}$ is the set of vertices representing the individuals and $\mathbb{E} \subseteq  \mathbb{V} \times \mathbb{V}$ is the set of edges of the digraph $\mathfrak{G}$  representing the influence structure. We assume that the social network has no self-loops, i.e., for any $\mathrm{v}_{\mathrm{i}} \in \mathbb{V}$, $(\mathrm{v}_{\mathrm{i}}, \mathrm{v}_{\mathrm{i}}) \notin \mathbb{E}$. The communication from information sources to individuals is modeled by a  bipartite digraph $\mathfrak{H} = (\mathbb{V} \bigcup \mathbb{K}, \mathbb{B})$, where $\mathbb{K}$ = $\left\{\mathrm{u}_{1}, \ldots,  \mathrm{u}_{\mathrm{f}}\right\}$ is the set of vertices representing the information sources and $\mathbb{B} \subset \mathbb{V} \times \mathbb{K}$ is the set of edges of the digraph. $\mathbb{I}$ denotes the set of followers of information sources, i.e., $\mathbb{I} = \left\{ {\left. {{\mathrm{v}_i}} \right|\left( {{\mathrm{v}_\mathrm{i}},{\mathrm{u}_\mathrm{d}}} \right) \in \mathbb{B},{\mathrm{v}_\mathrm{i}} \in \mathbb{V},{\mathrm{u}_\mathrm{d}} \in \mathbb{K}} \right\}$.

\subsection{Social Network Model}
In this paper, we use the opinion dynamics in \cite{mao2018spread}:
\begin{align}
\!\!\!\!\!{x_i}(k \!+\! 1) \!=\! {\alpha_i}(x_{i}(k)){s_i} + \sum\limits_{j \in \mathbb{V}} \!{{ w_{i,j}}{x_j}(k)}  \!+\!\! \sum\limits_{d \in \mathbb{K}} \!\!{{{\hat w}_{i,d}}(x_{i}(k)){u_d}}
\label{od1}
\end{align}
where
\begin{enumerate}
\item $x_i(k) \in [0,1]$ is individual $\mathrm{v}_{\mathrm{i}}$'s opinion at time $k$, $s_i$ is her fixed innate opinion, $u_{d} \in [0,1]$ is the information source $\mathrm{u}_{\mathrm{d}}$'s opinion;
\item $w_{i,j}$ represents the fixed weighted influence of individual $\mathrm{v}_{\mathrm{j}}$ on individual $\mathrm{v}_{\mathrm{i}}$,
  \begin{equation}
	w_{i,j}  \begin{cases}
		> 0, & \text{if } (\mathrm{v}_\mathrm{i},\mathrm{v}_\mathrm{j}) \in \mathbb{E}\\
		= 0, & \text{otherwise};
	\end{cases}\nonumber
  \end{equation}
  \item $\hat{w}_{i,d}(x_{i}(k))$ is the weighted influence of information source $\mathrm{u}_{\mathrm{d}}$ on individual $\mathrm{v}_{\mathrm{i}}$ with
  \begin{equation}
  {\hat w_{i,d}}({{x_i}(k)}) = \begin{cases}
		{g_{i,d}}({\left| {{x_i}(k) \!-\! {u_d}} \right|}), &\text{if} (\mathrm{v}_\mathrm{i},\mathrm{u}_\mathrm{d}) \in \mathbb{B}\\
		0, & \text{otherwise};\label{cb}
	\end{cases}
 \end{equation}
where ${g_{i,d}}\left(\cdot\right): \mathbb{R} \rightarrow \mathbb{R}$ is a strictly decreasing function that models confirmation bias: an individual tends to seek out, and consequently be influenced more by, an information source who reflects beliefs closer to hers. We assume that $g_{i,d}(\cdot)$ satisfies $1 > {g_{i,d}}\left( {\left| {{x_i}( k) - {u_d}} \right|} \right) > 0$.
  \item $\alpha_i(x_i(k))$ is the ``resistance parameter'' of individual $\mathrm{v}_{\mathrm{i}}$,  determined in such a way that it satisfies
      \begin{align}
      {\alpha _i}({{x_i}(k)}) + \sum\limits_{j \in \mathbb{V}} \!{{w_{i,j}} + \sum\limits_{d \in \mathbb{K}}\! {{{\hat w}_{i,d}}({{x_i}(k)})} } = 1, \,\, i \in\! \mathbb{V}.\label{convex}
      \end{align}
\end{enumerate}

\begin{rem}
We make a common assumption, similarly made in several related work, see e.g., ~\cite{das2013debiasing}, as the individual innate opinion is  regarded as her initial opinion, i.e.,
 \begin{align}
{s_i} = {x_i}(0),  \quad i \in \mathbb{V}. \label{inin}
\end{align}
\end{rem}

\subsection{Related Prior Work}

The most relevant prior work is the approach in \cite{WHT10}, where an exact inference procedure for \emph{undirected} network topology in conjunction with the continuous-time consensus dynamics is proposed. The discrete-time version of consensus dynamics considered therein is
\begin{align}
\tilde{\textbf{x}}(k+1) =  \widetilde{\textbf{L}} \tilde{\textbf{x}}(k), \label{dem1}
\end{align}
where $\widetilde{\textbf{L}} \in \mathbb{R}^{n \times n}$ is a symmetric row stochastic matrix. The exact inference procedure of $\widetilde{\textbf{L}}$ is based on the well-known numerical solutions of the constrained Lyapunov equation:
\begin{align}
\widetilde{\textbf{L}}\textbf{U} + \textbf{U}\widetilde{\textbf{L}} = \textbf{V}, ~~\widetilde{\textbf{L}} \in \mathbb{R}^{n \times n},~~\widetilde{\textbf{L}}\mathbf{1} = \mathbf{1},~~[\widetilde{\textbf{L}}]_{i,j} \geq 0 \label{dem2}
\end{align}
where $\mathbf{V} \triangleq \sum\limits_{k = 0}^{m - 1} {({\tilde{\mathbf{x}}({k + 1}){\tilde{\mathbf{x}}^\top}(k) + \tilde{\mathbf{x}}(k){\tilde{\mathbf{x}}^\top}({k + 1})})}$ and $\mathbf{U} \triangleq \sum\limits_{k = 0}^{m - 1} {\tilde{\mathbf{x}}(k)}{\tilde{\mathbf{x}}^\top}(k)$. However, this exact inference method works only for undirected communication topologies, i.e., this method cannot generate a unique inference solution for directed communication topologies when $n > 3$, as demonstrated as follows.

In the context of directed communication, i.e., $[\widetilde{\textbf{L}}]_{i,j} \neq [\widetilde{\textbf{L}}]_{j,i}$ for some $i \neq j \in \mathbb{V}$, the relation \eqref{dem2} would be
\begin{align}
\widetilde{\textbf{L}}\mathbf{U} + \mathbf{U}\widetilde{\textbf{L}}^{\top} = \mathbf{V}, ~~\widetilde{\textbf{L}} \in \mathbb{R}^{n \times n},~~\widetilde{\textbf{L}}\mathbf{1} = \mathbf{1},~~[\widetilde{\textbf{L}}]_{i,j} \geq 0. \label{dem3}
\end{align}

For the directed communication graph with $n$ agents, there are $n^{2} - n$ weighted communication links (variables) $[\widetilde{\textbf{L}}]_{i,j}$, $i \neq j$, that need to be inferred. In \eqref{dem3}, both the matrices $\widetilde{\textbf{L}}\textbf{U} + \textbf{U}\widetilde{\textbf{L}}^{\top}$ and $\textbf{V}$  are symmetric.  Thus, \eqref{dem3} contains, at most, $\frac{{\left( {n + 1} \right)n}}{2}$ distinct linear equations that are related to one or some $[\widetilde{\textbf{L}}]_{i,j}, i \neq j$. Moreover, we note that the constraint conditions $\widetilde{\textbf{L}}\mathbf{1} = \mathbf{1}$ and $[\widetilde{\textbf{L}}]_{i,j} \geq 0$ in \eqref{dem3} only reduce the number of linear equations while do not affect the number of variables to be inferred. This implies that if $({n^2} - n) - \frac{{\left( {n + 1} \right)n}}{2} = \frac{{\left( {n - 3} \right)n}}{2} > 0$ if $n > 3$, i.e., if the network includes more than three agents, then the number of unknown weights that need to be inferred is larger than the number of equations included in \eqref{dem3}. Hence, the procedure in \cite{WHT10} cannot be applied to the directed communication graph when $n > 3$.


\subsection{Problem Formulation}
In this paper, we  first investigate the setting where all information source opinions are under control  and confirmation bias follows a piece-wise model, i.e., the function ${g_{i,d}}\left(\cdot\right)$ in \eqref{cb} is described by
\begin{equation}
g_{i,d}( x_i(k)) = \beta_{i} - \gamma_{i}|x_i(k) - u_{d}|.\label{lmodel}
\end{equation}

 We next  study the case without confirmation bias and information source opinions are not controlled. We finally analyze perhaps the most realistic setting where the model of confirmation bias is unknown and the information sources are not under control. For this setting, we only infer an approximate topology. The studied problems in the three different cases are formally stated as follows.

\textbf{Inference Problem I}: For the opinion evolution {\bf with confirmation bias} ($\gamma_{i} \neq 0$ for some $i \in \mathbb{V}$), given {\bf controlled} opinions of information sources and measured evolving opinions $\textbf{x}(k)$ at some time, {\bf exactly} infer the network topology and the confirmation bias.

\textbf{Inference Problem II}: For the opinion evolution {\bf without confirmation bias} ($\gamma_{i} = 0$ for any $i \in \mathbb{V}$), given {\bf uncontrolled information sources} and measured evolving opinions $\textbf{x}(k)$ at some time, {\bf exactly} infer social network topology.

\textbf{Inference Problem III}: With {\bf unknown} confirmation bias model, given  {\bf uncontrolled opinions} of information sources and measured evolving opinions $\textbf{x}(k)$ at some time, {\bf approximately} infer social network topology.
%

\begin{rem}
In order to obtain the exact inference, we need a ``global capability": measure all of the individuals' evolving opinions for some time period. As also mentioned in Remark 2 of \cite{WHT10}, such a global capability is necessary  in the exact topology reconstruction with consensus-seeking dynamics. 
\end{rem}

\section{Inference Problem I}
 In this problem, information source opinions are control variables, hence, to simplify the inference procedure, we set them to ``zero": \begin{align}
u_{1} = u_{2} = \ldots = u_{m} = 0. \label{opii}
\end{align}

Since the information sources express the same opinion, we mathematically treat them as one information source $\mathrm{u}$. It follows from  \eqref{convex}
and  \eqref{od1} that $x_{i}(k) \in [0,1], \forall i \in \mathbb{V}, \forall k \in \mathbb{N}$, and hence:
\begin{equation}
	{\hat w_{i,d}}({{x_i}(k)}) = \beta_{i} - \gamma_{i}x_i(k).\label{stsdep}
  \end{equation}
We note here that the communication from information source $\mathrm{u}$ to individuals is incorporated into the parameters $\beta_{i}$ and $\gamma_{i}$:
\begin{align}
   \beta_{i}\,\,\,  \begin{cases}
		 > 0, \!\!\!\!&(\mathrm{v}_\mathrm{i},\mathrm{u}) \!\in\! \mathbb{B},\\
		=0, \!\!\!\!&\text{otherwise},
	\end{cases}, \quad \gamma_{i} \,\,\,\begin{cases}
		> 0, \!\!\!\!&(\mathrm{v}_\mathrm{i},\mathrm{u}) \!\in\! \mathbb{B},\\
		=0, \!\!\!\!&\text{otherwise};
	\end{cases}.\label{encd}
\end{align}
The resistance parameters are obtained from  \eqref{convex} as
\begin{align}
{\alpha _i}({x_i}(k)) = 1 - \sum\limits_{j \in \mathbb{V}}{w_{i,j}} - {\beta _i} + {\gamma _i}{x_i}(k),  \quad i \in \mathbb{V}.\label{respa}
\end{align}

Under the settings of \eqref{inin} and \eqref{opii}, it follows from \eqref{stsdep} and \eqref{respa} that  \eqref{od1} can equivalently be expressed as
\begin{align}
\textbf{x}(k + 1) = \textbf{A}\textbf{x}(0) + \textbf{W}\textbf{x}(k),\label{eq:hh1}
\end{align}
where we define:
\begin{subequations}
\begin{align}
& \textbf{x}(k) \triangleq [x_{1}(k), x_{2}(k), \ldots,  x_{n}(k)]^\top, \label{evs}\\
& \textbf{A} \triangleq \text{diag}\{{1 \!-\! \sum\limits_{j \in \mathbb{V}} {{w_{1,j}}}  \!-\! {\beta_1}, \ldots, 1 \!-\! \sum\limits_{j \in \mathbb{V}} {{w_{n,j}}}  \!-\! {\beta _n}} \},\label{eq:dmk}\\
&[\textbf{W}]_{i,j} \triangleq \begin{cases}
		{\gamma _i}{x_i}(0), &i = j \in \mathbb{V}\\
		w_{i,j}, &i \neq j \in \mathbb{V}.
	\end{cases}\label{eq:dm1}
\end{align}\label{orgpa}
\end{subequations}

\subsection{Solvability of Inference Problem I}
We now consider the following dynamics
\begin{align}
\textbf{x}(k + 1) = \widetilde{\textbf{A}}\textbf{x}(0) + \widetilde{\textbf{W}}\textbf{x}(k), \label{eq:hh2}
\end{align}
where $\widetilde{\textbf{A}}$ and $\widetilde{\textbf{W}}$ are inferred matrices of $\textbf{A}$ and $\textbf{W}$, respectively. These matrices are defined as
\begin{subequations}
\begin{align}
& \widetilde{\textbf{A}} \triangleq \text{diag}\{{1 \!-\! \sum\limits_{j \in \mathbb{V}} {{\tilde{w}_{1,j}}}  \!-\! {\tilde{\beta}_1}, \ldots, 1 \!-\! \sum\limits_{j \in \mathbb{V}} {{\tilde{w}_{n,j}}}  \!-\! {\tilde{\beta}_n}} \},\label{eq:dmka}\\
&[\widetilde{\textbf{W}}]_{i,j} \triangleq \begin{cases}
		{\tilde{\gamma}_i}{x_i}(0), &i = j \in \mathbb{V}\\
		\tilde{w}_{i,j}, &i \neq j \in \mathbb{V}.
	\end{cases}\label{eq:dm1a}
\end{align}\label{orgpaa}
\end{subequations}

Based on  \eqref{eq:hh1} and \eqref{eq:hh2}, we now define the solvability of exact inference problem.
\begin{defa}[Solvability]
Given the measurements of opinions $\textbf{x}(k)$, $k$ $=$ $0$, $1$, $\ldots$, $m$, the exact inference problem is said to be solvable if and only if two following two conditions are satisfied:
\begin{description}
  \item[C1:] $\textbf{x}(k)$ is evolving according to both \eqref{eq:hh1} and \eqref{eq:hh2} for time $k = 0,1, \ldots, m$;
  \item[C2:] $\widetilde{\textbf{W}}$ $=$ ${\textbf{W}}$, $\tilde{\beta}_{i}$ $=$ $\beta_{i}$ and $\tilde{\gamma}_{i}$ $=$ $\gamma_{i}$, $i$ $\in$ $\mathbb{V}$.
\end{description}
\end{defa}

\begin{rem}
We note that if C2 had required only $\widetilde{\textbf{A}} = \textbf{A}$ and/or $\widetilde{\textbf{W}} = \textbf{W}$, by \eqref{eq:dm1} we could only infer the network topology: $w_{i,j} = [\widetilde{\textbf{W}}]_{i,j}, i \neq j \in \mathbb{V}$. However, the inference of  bias parameters would not be unique, as implied by:
$[\widetilde{\textbf{W}}]_{i,i} = {\tilde{\gamma }_i}{x_i}(0) = {{\gamma }_i}{x_i}(0) = [\textbf{W}]_{i,i}$, $i \in \mathbb{V}$, when $x_{i}(0) = 0$.\label{rem3}
\end{rem}

We next continue analyzing the conditions of solvability, which paves the way for derivations of (exact and approximated) inference procedures.

\begin{prop}
$\textbf{x}(k)$ is evolving according to both  \eqref{eq:hh1} and   \eqref{eq:hh2} for $k = 1, \ldots, m$,  if and only if
\begin{align}
\textbf{x}(0) \in {\textbf{L}^{ - 1}}\ker({\widehat{\textbf{O}}}) \cap \ker ({\widehat{\textbf{A}} + \widehat{\textbf{W}}} ),\label{iffc}
\end{align}
where
\begin{subequations}
\begin{align}
\textbf{L} &\triangleq \textbf{A} + \textbf{W} - \textbf{I}, \label{defla}\\
\widehat{\textbf{A}} &\triangleq \widetilde{\textbf{A}} - \textbf{A},~~~\widehat{\textbf{W}} \triangleq \widetilde{\textbf{W}} - \textbf{W}, \label{defma}\\
\widehat{\textbf{O}} &\triangleq {[{\widehat{\textbf{W}}^\top, ({\widehat{\textbf{W}}\textbf{W}})^\top, \ldots , (\widehat{\textbf{W}}{\textbf{W}^{m-1}}})^\top]^\top}.\label{defmacc}
\end{align}\label{iff}
\end{subequations}\label{prop1}
\end{prop}

\begin{IEEEproof}
See Appendix A.
\end{IEEEproof}

Based on Proposition \ref{prop1}, we directly obtain the necessary and sufficient condition on the solvability of the inference problem.
\begin{cor}
Consider the social dynamics \eqref{eq:hh1} with \eqref{orgpa}, and \eqref{eq:hh2} with \eqref{orgpaa}. The inference of network topology and confirmation bias is solvable for \eqref{eq:hh1}, if and only if
\begin{align}
\textbf{x}(0) \notin  {\textbf{L}^{ - 1}}\ker({\widehat{\textbf{O}}}) \cap \ker ({\widehat{\textbf{A}} + \widehat{\textbf{W}}} )\nonumber
\end{align}
holds for any $w_{i,j} \neq \tilde{w}_{i,j}, i \neq j \in \mathbb{V}$, or $\beta_{i} \neq \tilde{\beta}_{i}, i \in \mathbb{V}$, or $\gamma_{i} \neq \tilde{\gamma}_{i}, i \in \mathbb{V}$.
\label{cor1}
\end{cor}

We note that Corollary \ref{cor1} requires the knowledge of unavailable inference errors of encoded matrices, i.e., $\widehat{\textbf{A}}$ and $\widehat{\textbf{W}}$. In the following, we provide a sufficient condition which does not require these matrices.

\begin{thm}
The exact inference problem for the social dynamics \eqref{eq:hh1} is solvable if
\begin{align}
&{x_i}( 0 ) \ne 0,~\text{for}~\forall (\mathrm{v}_{i}, \mathrm{u}) \in \mathbb{B},\label{suff2}\\
&\text{rank}({[{\textbf{L}\textbf{x}(0),\textbf{W}\textbf{L}\textbf{x}(0), \ldots ,{\textbf{W}^{n - 1}}\textbf{L}\textbf{x}(0)}]}) = n. \label{suff1}
\end{align}\label{thm1}
\end{thm}

\begin{IEEEproof}
See Appendix B.
\end{IEEEproof}

\subsection{Exact Solution to Inference Problem I}
We note that while (\ref{suff2}) and (\ref{suff1}) guarantee a unique inference solution, in practice we cannot use these to device an algorithm to solve exact inference problem,  since the matrices $\textbf{W}$ and $\textbf{L}$ are unavailable. Towards designing a practical algorithm, we now consider the following measurement matrix:
\begin{align}
\textbf{P} \triangleq \sum\limits_{k = 0}^{m - 1} {({\textbf{x}(k + 1) \!-\! \textbf{x}(k)})} {({\textbf{x}(k + 1) \!-\! \textbf{x}(k)})^\top}\!,m \in \mathbb{N}.\label{defmatb}
\end{align}


The following auxiliary lemmas, whose proofs appear in Appendices C and D, we present  properties of $\textbf{P}$ which will be used in the derivation of the proposed inference procedures.
\begin{lem}
Consider the matrix \eqref{defmatb}. For the social dynamics \eqref{eq:hh1}, we have
\begin{align}
\textbf{W}\textbf{P} = \textbf{Q}, \label{lyapunovb}
\end{align}
where
\begin{align}
\textbf{Q} &\triangleq \sum\limits_{k = 0}^{m - 1} {({\textbf{x}(k + 2) - \textbf{x}(k + 1)}){{({\textbf{x}(k + 1) - \textbf{x}(k)})^\top}}}.\label{defly}
\end{align}
\label{lem1}
\end{lem}

\begin{lem}
Consider the matrix \eqref{defmatb}.  For the social dynamics \eqref{eq:hh1}, we have
\begin{align}
\ker (\textbf{P}) = \ker ([{\textbf{L}\textbf{x}(0),\textbf{W}\textbf{L}\textbf{x}(0), \ldots ,{\textbf{W}^{m - 1}}\textbf{L}\textbf{x}(0)} ]^\top), \label{reslem2}
\end{align}
where $\textbf{L}$ is given by \eqref{defla}. \label{lem2}
\end{lem}

In  \eqref{lyapunovb},  $\textbf{P}$ and $\textbf{Q}$ are known since they are computed from the available measurements $\textbf{x}(k)$, $k = 0, 1, m-1$. If the encoded matrix $\textbf{W}$ can be uniquely obtained from \eqref{lyapunovb}, we can uniquely infer network influence weights $w_{i,j} = {\left[ \textbf{W} \right]_{i,j}}$, with $i \neq j$, which describe the network topology ($(\mathrm{v}_{i}, \mathrm{v}_{j}) \in \mathbb{E}$ if $w_{i,j} \neq 0$), and the bias parameters ${\gamma _i} = \frac{{{{\left[ \textbf{W} \right]}_{i,i}}}}{{{x_i}( 0 )}}$ that contain the information of communication from information source $\mathrm{u}$ to individuals
($(\mathrm{v}_{i}, \mathrm{u}) \in \mathbb{B}$ if $\gamma_{i} \neq 0$). With the obtained $\mathbf W$, the remaining parameters $\beta_{i}$ can be obtained as follows.

\begin{cor}
Consider social dynamics \eqref{eq:hh1}. Given $\textbf{W}$, $\textbf{x}(0)$, $\textbf{x}(k)$ and $\textbf{x}(k+1)$, the parameters $\beta_{i}$ of  bias are :
\begin{align}
{\beta _i} \!=\! 1 \!-\! \!\!\sum\limits_{j \ne i \in \mathbb{V}} \!\!{{{[\textbf{W}]}_{i,j}}}  \!-\! \frac{{ {x_i}({k \!+\! 1}) \!-\! \sum\limits_{j \in \mathbb{V}} {{{[\textbf{W}]}_{i,j}}} {x_j}(k)}}{{{x_i}(0)}}, i \!\in\! \mathbb{V}. \label{confirp}
\end{align}
\label{cor2}
\end{cor}

\begin{IEEEproof}
See Appendix E.
\end{IEEEproof}

In the following theorem, whose proof is presented in Appendix F, we present our results on the exact inference problem.
\begin{thm}
The exact inference problem is solvable for  \eqref{eq:hh1}, if and only if \eqref{suff2} holds and there exists a unique $\widetilde{\textbf{W}} \in \mathbb{R}^{n \times n}$ such that
\begin{align}
\widetilde{\textbf{W}} \textbf{P} = \textbf{Q}, \label{thm21}
\end{align}
where $\textbf{P}$ and $\textbf{Q}$ are defined in \eqref{defmatb} and \eqref{defly}, respectively. \label{thm2}
\end{thm}

The solvability of inference problem should be checked before the computation \eqref{thm21}. However, the solvability condition presented in Theorem~\ref{thm1} requires the knowledge of $\textbf{L} = \textbf{W} + \textbf{A} - \textbf{I}$ that is unavailable, which makes this result unusable in practice. As a remedy,   in the following theorem (whose proof is presented in Appendix G), we present a new method that  makes use of the available $\textbf{P}$ instead of $\textbf{L}$ to check the solvability, and then solve the inference problem.
\begin{thm}
Consider the matrices $\textbf{P}$ and $\textbf{Q}$ given by \eqref{defmatb} and \eqref{defly}, respectively. If the condition \eqref{suff2} holds and $\text{rank}(\textbf{P}) = n$, the network topology and bias parameters are exactly inferred as
\begin{align}
&{w_{i,j}} = {[\textbf{Q}{\textbf{P}^{ - 1}}]_{i,j}}, ~~i \ne j \in \mathbb{V} \label{thm31}\\
& {\gamma _i} = \frac{{{{[\textbf{Q}{\textbf{P}^{ - 1}}]}_{i,i}}}}{{{x_i}\left( 0 \right)}}, ~~~i \in \mathbb{V}
 \label{thm32} \\
&\!\!\!\!\!{\beta _i} = 1 - \sum\limits_{j \ne i \in \mathbb{V}} {{{[\textbf{Q}{\textbf{P}^{ - 1}}]}_{i,j}}} \nonumber\\
&\hspace{0.70cm} - ( {{x_i}(k + 1) - \sum\limits_{j \in \mathbb{V}} {{{[\textbf{Q}{\textbf{P}^{ - 1}}]}_{i,j}}} {x_j}(k)})\frac{1}{{{x_i}(0)}}, i \!\in\! \mathbb{V}.
\label{thm33}
\end{align}
\label{thm3}
\end{thm}

\begin{rem}
Using the inferred matrices $\textbf{W}$ and $\textbf{A}$ and the available data $\textbf{x}(0)$, the steady state of evolving opinions is exactly inferred from $\textbf{A}\textbf{x}(0) + \textbf{W}\textbf{x}^* = \textbf{x}^*$:
\begin{align}
{\textbf{x}^*} = {({\textbf{I} - \textbf{W}})^{ - 1}}\textbf{A}\textbf{x}(0).\nonumber
\end{align}
\end{rem}

\section{Inference Problem II}
In this section, we consider the scenario with no confirmation bias, which is  described by \eqref{lmodel} with $\gamma_{i} = 0, \forall i \in \mathbb{V}$. If the opinions of information sources are still controlled, \eqref{thm31} and \eqref{thm33}  generate the exact topology inference, hence the problem is trivial. Therefore, in this scenario, we investigate \emph{whether the topology can still be exactly inferred even if the opinions of information sources are not under control but still known to the inference algorithm designer}.

We  consider the dynamics that is slightly modified from  \eqref{od1}:
\begin{align}
{x_i}(k \!+\! 1) = {\alpha _i}{s_i} + \sum\limits_{j \in \mathbb{V}} {{w_{i,j}}} {x_j}(k) + \sum\limits_{d \in \mathbb{K}} {{{\hat w}_{i,d}}} {u_d}, i \!\in\! \mathbb{V} \label{od1ncb}
\end{align}
where $\hat{w}_{i,d}$ represents the fixed weighted influence of information source $\mathrm{u}_{\mathrm{d}}$ on individual $\mathrm{v}_{\mathrm{i}}$, the fixed resistance parameter $\alpha_i$ of individual $\mathrm{v}_{\mathrm{i}}$ is determined in such a way that it satisfies
 \begin{align}
{\alpha _i} + \sum\limits_{j \in \mathbb{V}}  {w_{i,j}} + \sum\limits_{d \in \mathbb{K}} {{{\hat w}_{i,d}}}  = 1, \forall i \in \mathbb{\mathbb{V}}.\label{convexncb}
 \end{align}

Under the setting of \eqref{inin}, it follows from  \eqref{convexncb} that \eqref{od1ncb} can be equivalently expressed as the following.
\begin{align}
\mathbf x(k + 1) = \mathcal{A}\textbf{x}(0) + \mathcal{W}\mathbf x(k),\label{eq:hh1ncb}
\end{align}
where we define
\begin{align}
&\mathcal{A} \triangleq \text{diag}\{1 - \sum\limits_{j \in \mathbb{V}} {{w_{1,j}}}  - \sum\limits_{d \in \mathbb{K}} {{{\hat w}_{1,d}}}  + \sum\limits_{d \in \mathbb{K}} {\frac{{{{\hat w}_{1,d}}{u_d}}}{{{x_1}(0)}}} , \ldots , \nonumber\\
&\hspace{1.5cm}1 - \sum\limits_{j \in \mathbb{V}} {{w_{n,j}}}  - \sum\limits_{d \in \mathbb{K}} {{{\hat w}_{n,d}}}  + \sum\limits_{d \in \mathbb{K}} {\frac{{{{\hat w}_{n,d}}{u_d}}}{{{x_n}(0)}}}\},\label{eq:dmkncb}\\
&\left[ \mathcal{W} \right]_{i,j} \triangleq \begin{cases}
		0, &i = j \in \mathbb{V}\\
		w_{i,j}, &i \neq j \in \mathbb{V}.
	\end{cases}\label{eq:dm1ncb}
\end{align}

We note that the dynamics \eqref{eq:hh1} and \eqref{eq:hh1ncb} have the same form. Therefore, the analysis method in deriving the inference procedure for  \eqref{eq:hh1} can be employed for  \eqref{eq:hh1ncb}.

\begin{cor}
Consider the matrices $\textbf{P}$ and $\textbf{Q}$ given by \eqref{defmatb} and \eqref{defly}, respectively. If $\text{rank}(\textbf{P}) = n$, network topology associated with dynamics \eqref{eq:hh1ncb} is exactly inferred as
\begin{align}
\mathcal{W} = \textbf{Q}{\textbf{P}^{ - 1}}. \label{thm31ncb}
\end{align}
\label{thm3}
\end{cor}
\begin{rem}We obtain from  \eqref{eq:hh1ncb} that
\begin{align}
&\sum\limits_{d \in \mathbb{K}} {{{\hat w}_{i,d}}} ({{u_d} - {x_i}(0)}) \nonumber\\
& = {x_i}({k \!+\! 1}) \!-\! \sum\limits_{j \in \mathbb{V}} {{{[\mathcal{W}]}_{i,j}}{x_j}(k)}  \!-\! ({1 \!-\! \sum\limits_{j \in \mathbb{V}} {{{[\mathcal{W}]}_{i,j}}} }){x_i}(0), \label{singal}
\end{align}
whose right-hand side is known, since the evolving and innate opinions ${x_i}(k)$ and ${x_i}(0)$, $i \in \mathbb{V}$, are available measurement data, and $\mathcal{W}$ is obtained from \eqref{thm31ncb}. When the network has multiple information sources, i.e.,  $\left| \mathbb{K} \right| \ge 2$, the left-hand side of \eqref{singal} has more than one variables ${\hat w}_{i,d}, d \in \mathbb{K}$, to be inferred from one equation. In this case, topology cannot be uniquely inferred without controlling information sources.\end{rem}

\section{Inference Problem III}
In Problem I, the solution is based on the assumption that the bias function follows a piece-wise linear model. In this section, we explore what can be obtained when this assumption is removed. We show that Theorem \ref{thm3} can still be used to approximate the inference solution, however an exact inference of network topology cannot be obtained.

Under \eqref{inin}, following \eqref{convex},  we re-express  \eqref{od1}:
\begin{align}
\textbf{x}(k + 1) = \breve{\textbf{A}}(k)\textbf{x}(0) + \breve{\textbf{W}}\textbf{x}(k),\label{nsd}
\end{align}
where we define:
\begin{align}
\!\!\!\!&\breve{\textbf{A}}(k) \!\triangleq\! \text{diag} \{1 \!-\! \sum\limits_{j \in \mathbb{V}}\! {{\breve{w}_{1,j}}}  \!-\! \sum\limits_{d \in \mathbb{K}}\!{{{\breve{w}}_{1,d}}({{{x}_1}(k)})}  \!+\!\! {\frac{{{{\breve{w}}_{1,d}}({{{x}_1}(k)} ){u_d}}}{{{{x}_1}(0)}}}, \nonumber\\
&\hspace{0.0cm}\ldots, 1 \!-\!\! \sum\limits_{j \in \mathbb{V}}\! {{\breve{w}_{n,j}}}  \!-\!\! \sum\limits_{d \in \mathbb{K}}\!{{{\breve{w}}_{n,d}}({{{x}_n}(k)})}  \!+\! {\frac{{{{\breve{w}}_{n,d}}({{{x}_n}(k)} ){u_d}}}{{{{x}_n}(0)}}}\}\!,\label{eq:ks1}\\
\!\!\!\!&[\breve{\textbf{W}}]_{i,j} \triangleq \begin{cases}
		0, &i = j \in \mathbb{V}\\
		\breve{w}_{i,j}, &i \neq j \in \mathbb{V}.
	\end{cases}\label{eq:ks2}
\end{align}

We next consider the following symmetric matrix:
\begin{align}
\!\!\breve{\textbf{P}}_{m,p} \!\triangleq\! \sum\limits_{k = m}^{p} \!{({{\textbf{x}}(k \!+\! 1) \!-\! {\textbf{x}}(k)})} {({{\textbf{x}}(k \!+\! 1) \!-\! {\textbf{x}}(k)})^\top}\!,  m \!\in\! \mathbb{N}.\label{noidefm}
\end{align}

\begin{lem}
Consider the matrix \eqref{noidefm}.  For \eqref{nsd}, we have
\begin{align}
\breve{\textbf{W}}\breve{\textbf{P}}_{m,p} = \breve{\textbf{Q}}_{m,p} + \breve{\textbf{R}}_{m,p}, \label{lm3r}
\end{align}
where
\begin{align}
\breve{\textbf{Q}}_{m,p} \triangleq &\sum\limits_{k = m}^{p} \!\!{({{\textbf{x}}(k \!+\! 2) \!-\! {\textbf{x}}(k \!+\! 1)}){{({{\textbf{x}}(k \!+\! 1) \!-\! {\textbf{x}}(k)})}^\top}},\label{lm3rd1} \\
\breve{\textbf{R}}_{m,p} \triangleq &\sum\limits_{k = m}^{p} \!\!{({\breve{\textbf{A}}({k \!+\! 1}) \!-\! \breve{\textbf{A}}(k)})\textbf{x}(0){{({{\textbf{x}}(k \!+\! 1) \!-\! {\textbf{x}}(k)})}^\top}}. \label{lm3rd2}
\end{align}
\label{lem3}
\end{lem}

\begin{IEEEproof}
See Appendix H.
\end{IEEEproof}

The key idea here is the following. Due to the unknown model of confirmation bias, $\breve{\textbf{A}}({k + 1}) - \breve{\textbf{A}}(k)$ is unavailable. Therefore, \eqref{lm3r} cannot be used for exact inference. However, \eqref{lm3r} with $\mathbf P$ replaced by $\breve{\mathbf{P}}_{m,p}$ can be used to {\it exactly} infer {\it partial} network topology.  This idea is stated formally in the following theorem.

\begin{thm}
Consider the social dynamics \eqref{nsd} with unknown confirmation bias model. If an individual is not a follower of information sources, its weighted communication topology from its neighbors can be exactly inferred from
\begin{align}
\textbf{W}\breve{\textbf{P}}_{m,p} = \breve{\textbf{Q}}_{m,p}, \text{with}~\text{rank}(\breve{\textbf{P}}_{m,p}) = n, \label{app}
\end{align}
and we have
\begin{align}
{\breve{w}_{i,j}} = {w}_{i,j} =  {[{\breve{\textbf{Q}}_{m,p}{\breve{\textbf{P}}^{ - 1}_{m,p}}}]_{i,j}},~\text{if}~i \notin \mathbb{I}, j \in \mathbb{N}_{i}, \label{paifer}
\end{align}
where $\breve{\textbf{P}}_{m,p}$ and $\breve{\textbf{Q}}_{m,p}$ are given by \eqref{noidefm} and \eqref{lm3rd1}, respectively.
\label{thm4}
\end{thm}

\begin{IEEEproof}
See Appendix I.
\end{IEEEproof}

Finally, based on Theorem \ref{thm4}, we propose Algorithm~1 that generates approximate network topology.

\begin{algorithm}[http]
  \caption{Inference Algorithm for Problem III}
  \KwIn{Numbers $m \in \mathbb{N}$ and $p \geq m + n \in \mathbb{N}$, set
  \begin{align}
  \mathbb{P} = \{ {\left. k \right| \text{rank}({\textbf{P}_{k,p}}) = n, k = 1, \ldots, m} \},\label{set2}
  \end{align}
  where $\textbf{P}_{k,p}$ is defined by \eqref{noidefm}.}
    \eIf{$[\textbf{Q}_{r,p}{\textbf{P}_{r,p}^{ - 1}} - \textbf{Q}_{q,p}{\textbf{P}_{q,p}^{ - 1}}]_{i,j} = 0$, for $\forall r \neq q \in \mathbb{P}$}
    {
   $i \notin \mathbb{I}$ and ${\breve{w}_{i,j}} \!\leftarrow\! \begin{cases}
		0, &[\breve{\textbf{Q}}_{r,p}{\breve{\textbf{P}}_{r,p}^{ - 1}}]_{i,j} \!<\! 0\\
		[\breve{\textbf{Q}}_{r,p}{\breve{\textbf{P}}^{ - 1}}_{r,p}]_{i,j}, &\text{otherwise}.
	\end{cases}$ \;
    }
    {$i \in \mathbb{I}$ and ${\breve{w}_{i,j}} \!\leftarrow\! \begin{cases}
		0, &[\breve{\textbf{Q}}_{r,p}{\breve{\textbf{P}}_{r,p}^{ - 1}}]_{i,j} \!<\! 0\\
		[\breve{\textbf{Q}}_{r,p}{\breve{\textbf{P}}_{r,p}^{ - 1}}]_{i,j}, &\text{otherwise}.
	\end{cases}$;}
\end{algorithm}

\begin{rem}
We note that increasing the length of the observed data of evolving opinions does not necessarily yield smaller approximation error for this approximation. In fact, in our analysis, we have observed the opposite after some critical length. One intuitive explanation for this observation is that: as the length increases, $\textbf{x}(k)$ approaches the steady state, and consequently, $\textbf{x}(k)$ and $\textbf{x}(k + 1)$ are more correlated and $\breve{\textbf{P}}_{m,p}$ is more likely to be not full-rank.
\label{rem5}
\end{rem}

\section{Simulations}
In this section, we  start with the exact topology inference of an example network, we then study the approximate topology  inference in the context of a real social network which is known as the Krackhardt's advice network~\cite{krackhardt1987cognitive}.
\vspace{-0.2cm}
\subsection{Exact Inference}
\vspace{-0.2cm}
\begin{figure}[http]
\vspace{-0.3cm}
\centering{
\includegraphics[scale=0.45]{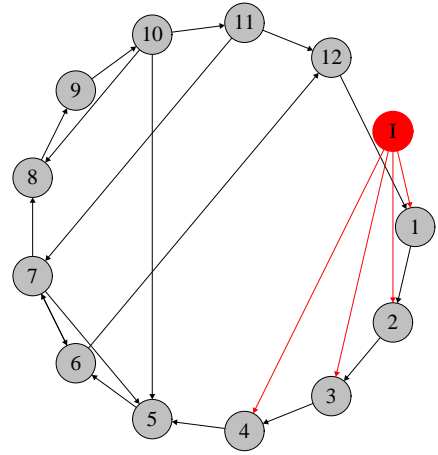}
}
\caption{Twelve individuals with one information source $\emph{\emph{I}}$.}
\label{rea1}
\end{figure}

\begin{figure}[http]
\centering{
\includegraphics[scale=0.37]{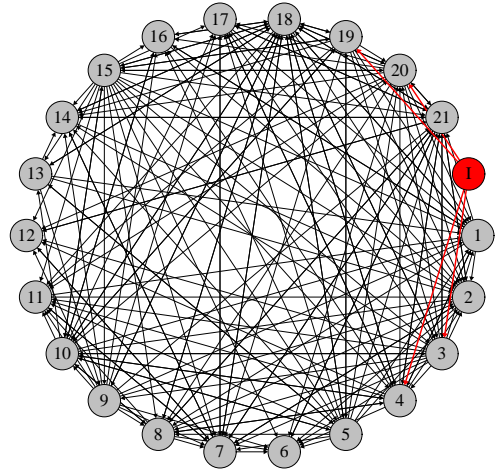}
}
\caption{Krackhardt's advice network~\cite{krackhardt1987cognitive} in the presence of one information source $\emph{\emph{I}}$.}
\label{mm}
\end{figure}

We consider the network with  $n = 12$ individuals in Figure~\ref{rea1}. The innate opinions are randomly generated as $\textbf{x}(0)$ $=$ $[{\rm{0}}{\rm{.7513,0}}{\rm{.2551,0}}{\rm{.506,0}}{\rm{.6991,0}}{\rm{.8909,0}}{\rm{.9593,0}}{\rm{.5472,0}}{\rm{.1386,}}$ \\ ${\rm{0}}{\rm{.1493,0}}{\rm{.2575,0}}{\rm{.8407,0}}{\rm{.2543}}]^{\top}$.
We let the information source $\emph{\emph{I}}$ express opinion $u_{d} = 0$. Then, following the model \eqref{lmodel}, we describe the confirmation biases  of the four followers of $\emph{\emph{I}}$ by
\begin{align}
g_{1,d}(x_1(k)) = 0.5 - 0.3x_1(k),\label{lmodl}\\
g_{2,d}(x_2(k)) = 0.4 - 0.2x_2(k),\label{lmod2}\\
g_{3,d}(x_3(k)) = 0.3 - 0.1x_3(k),\label{lmod3}\\
g_{4,d}(x_4(k)) = 0.2 - 0.1x_4(k).\label{lmod4}
\end{align}
\begin{figure*}[http]
\begin{align}\small
\textbf{P} &\!=\! \left[\!\!\! {\begin{array}{*{21}{c}}
0.1816  \!\!\!&\!\!\! -0.0818  \!\!\!&\!\!\!  0.1296  \!\!\!&\!\!\!  0.0725  \!\!\!&\!\!\!  0.1185  \!\!\!&\!\!\!  0.0628 \!\!\!&\!\!\!  -0.1209 \!\!\!&\!\!\!  -0.0511   \!\!\!&\!\!\! 0.0132  \!\!\!&\!\!\!  0.0207  \!\!\!&\!\!\!  0.2066 \!\!\!&\!\!\!  -0.1974  \\
-0.0818 \!\!\!&\!\!\!   0.0667  \!\!\!&\!\!\! -0.0634 \!\!\!&\!\!\!   0.0097 \!\!\!&\!\!\!  -0.0488 \!\!\!&\!\!\!  -0.0274  \!\!\!&\!\!\!  0.0770  \!\!\!&\!\!\!  0.0211  \!\!\!&\!\!\! -0.0209  \!\!\!&\!\!\! -0.0070 \!\!\!&\!\!\!  -0.0745  \!\!\!&\!\!\!  0.1248 \\
0.1296 \!\!\!&\!\!\!  -0.0634 \!\!\!&\!\!\!   0.0970 \!\!\!&\!\!\!   0.0399  \!\!\!&\!\!\!  0.0850  \!\!\!&\!\!\!  0.0455  \!\!\!&\!\!\! -0.0879  \!\!\!&\!\!\! -0.0327 \!\!\!&\!\!\!   0.0112  \!\!\!&\!\!\!  0.0108  \!\!\!&\!\!\!  0.1420 \!\!\!&\!\!\!  -0.1418\\
0.0725  \!\!\!&\!\!\!  0.0097 \!\!\!&\!\!\!   0.0399   \!\!\!&\!\!\! 0.1031 \!\!\!&\!\!\!   0.0564 \!\!\!&\!\!\!   0.0248  \!\!\!&\!\!\! -0.0214 \!\!\!&\!\!\!  -0.0282  \!\!\!&\!\!\! -0.0180  \!\!\!&\!\!\!  0.0186 \!\!\!&\!\!\!   0.1266 \!\!\!&\!\!\!  -0.0373\\
0.1185 \!\!\!&\!\!\!  -0.0488 \!\!\!&\!\!\!   0.0850 \!\!\!&\!\!\!   0.0564  \!\!\!&\!\!\!  0.0814  \!\!\!&\!\!\!  0.0411  \!\!\!&\!\!\! -0.0761 \!\!\!&\!\!\!  -0.0324  \!\!\!&\!\!\!  0.0034  \!\!\!&\!\!\!  0.0145  \!\!\!&\!\!\!  0.1469 \!\!\!&\!\!\!  -0.1231\\
0.0628  \!\!\!&\!\!\! -0.0274  \!\!\!&\!\!\!  0.0455  \!\!\!&\!\!\!  0.0248  \!\!\!&\!\!\!  0.0411  \!\!\!&\!\!\!  0.0221  \!\!\!&\!\!\! -0.0402 \!\!\!&\!\!\!  -0.0169  \!\!\!&\!\!\!  0.0044  \!\!\!&\!\!\!  0.0059  \!\!\!&\!\!\!  0.0703 \!\!\!&\!\!\!  -0.0652\\
-0.1209  \!\!\!&\!\!\!  0.0770 \!\!\!&\!\!\!  -0.0879  \!\!\!&\!\!\! -0.0214 \!\!\!&\!\!\!  -0.0761 \!\!\!&\!\!\!  -0.0402  \!\!\!&\!\!\!  0.1010  \!\!\!&\!\!\!  0.0350 \!\!\!&\!\!\!  -0.0188  \!\!\!&\!\!\! -0.0162 \!\!\!&\!\!\!  -0.1303  \!\!\!&\!\!\!  0.1658\\
-0.0511  \!\!\!&\!\!\!  0.0211 \!\!\!&\!\!\!  -0.0327 \!\!\!&\!\!\!  -0.0282  \!\!\!&\!\!\! -0.0324 \!\!\!&\!\!\!  -0.0169  \!\!\!&\!\!\!  0.0350 \!\!\!&\!\!\!   0.0183 \!\!\!&\!\!\!  -0.0037  \!\!\!&\!\!\! -0.0097 \!\!\!&\!\!\!  -0.0613  \!\!\!&\!\!\!  0.0591\\
0.0132  \!\!\!&\!\!\! -0.0209  \!\!\!&\!\!\!  0.0112 \!\!\!&\!\!\!  -0.0180 \!\!\!&\!\!\!   0.0034  \!\!\!&\!\!\!  0.0044  \!\!\!&\!\!\!  -0.0188  \!\!\!&\!\!\! -0.0037  \!\!\!&\!\!\!  0.0118 \!\!\!&\!\!\!  -0.0012 \!\!\!&\!\!\!  -0.0025 \!\!\!&\!\!\!  -0.0309\\
0.0207 \!\!\!&\!\!\!  -0.0070  \!\!\!&\!\!\!  0.0108  \!\!\!&\!\!\!  0.0186  \!\!\!&\!\!\!  0.0145  \!\!\!&\!\!\!  0.0059 \!\!\!&\!\!\!  -0.0162 \!\!\!&\!\!\!  -0.0097 \!\!\!&\!\!\!  -0.0012  \!\!\!&\!\!\!  0.0085 \!\!\!&\!\!\!   0.0334 \!\!\!&\!\!\!  -0.0284\\
0.2066 \!\!\!&\!\!\!  -0.0745   \!\!\!&\!\!\! 0.1420  \!\!\!&\!\!\!  0.1266  \!\!\!&\!\!\!  0.1469  \!\!\!&\!\!\!  0.0703 \!\!\!&\!\!\!  -0.1303  \!\!\!&\!\!\! -0.0613  \!\!\!&\!\!\! -0.0025  \!\!\!&\!\!\!  0.0334  \!\!\!&\!\!\!  0.2823  \!\!\!&\!\!\! -0.2118\\
-0.1974  \!\!\!&\!\!\!  0.1248 \!\!\!&\!\!\!  -0.1418  \!\!\!&\!\!\! -0.0373  \!\!\!&\!\!\! -0.1231  \!\!\!&\!\!\! -0.0652  \!\!\!&\!\!\!  0.1658  \!\!\!&\!\!\!  0.0591 \!\!\!&\!\!\!  -0.0309 \!\!\!&\!\!\!  -0.0284  \!\!\!&\!\!\! -0.2118  \!\!\!&\!\!\!  0.2741
\end{array}} \!\!\!\right],\label{dama1} \\
\textbf{Q} &\!=\! \left[\!\!\! {\begin{array}{*{21}{c}}
-0.0380  \!\!\!&\!\!\!  0.0315 \!\!\!&\!\!\!  -0.0275 \!\!\!&\!\!\!   0.0014 \!\!\!&\!\!\!  -0.0225 \!\!\!&\!\!\!  -0.0119  \!\!\!&\!\!\!  0.0391 \!\!\!&\!\!\!   0.0121 \!\!\!&\!\!\!  -0.0094 \!\!\!&\!\!\!  -0.0067 \!\!\!&\!\!\!  -0.0382 \!\!\!&\!\!\!   0.0651\\
0.0866  \!\!\!&\!\!\! -0.0375  \!\!\!&\!\!\!  0.0616  \!\!\!&\!\!\!  0.0367  \!\!\!&\!\!\!  0.0568  \!\!\!&\!\!\!  0.0300  \!\!\!&\!\!\! -0.0565 \!\!\!&\!\!\!  -0.0245 \!\!\!&\!\!\!   0.0055  \!\!\!&\!\!\!  0.0100  \!\!\!&\!\!\!  0.0995  \!\!\!&\!\!\! -0.0923\\
-0.0425  \!\!\!&\!\!\!  0.0368 \!\!\!&\!\!\!  -0.0331  \!\!\!&\!\!\!  0.0078 \!\!\!&\!\!\!  -0.0250 \!\!\!&\!\!\!  -0.0142 \!\!\!&\!\!\!   0.0418  \!\!\!&\!\!\!  0.0110  \!\!\!&\!\!\! -0.0120 \!\!\!&\!\!\!  -0.0037 \!\!\!&\!\!\!  -0.0375 \!\!\!&\!\!\!   0.0677\\
0.0958  \!\!\!&\!\!\! -0.0437  \!\!\!&\!\!\!  0.0707 \!\!\!&\!\!\!   0.0352 \!\!\!&\!\!\!   0.0634 \!\!\!&\!\!\!   0.0336  \!\!\!&\!\!\! -0.0630 \!\!\!&\!\!\!  -0.0249 \!\!\!&\!\!\!   0.0066  \!\!\!&\!\!\!  0.0088 \!\!\!&\!\!\!   0.1082  \!\!\!&\!\!\! -0.1018\\
-0.0107  \!\!\!&\!\!\!  0.0143 \!\!\!&\!\!\!  -0.0103 \!\!\!&\!\!\!   0.0116  \!\!\!&\!\!\! -0.0052  \!\!\!&\!\!\! -0.0038 \!\!\!&\!\!\!   0.0132  \!\!\!&\!\!\!  0.0013 \!\!\!&\!\!\!  -0.0059 \!\!\!&\!\!\!   0.0012 \!\!\!&\!\!\!  -0.0034  \!\!\!&\!\!\!  0.0209\\
-0.0125  \!\!\!&\!\!\!  0.0133 \!\!\!&\!\!\!  -0.0094  \!\!\!&\!\!\!  0.0049  \!\!\!&\!\!\! -0.0066 \!\!\!&\!\!\!  -0.0038  \!\!\!&\!\!\!  0.0151  \!\!\!&\!\!\!  0.0040 \!\!\!&\!\!\!  -0.0050  \!\!\!&\!\!\! -0.0020  \!\!\!&\!\!\! -0.0097  \!\!\!&\!\!\!  0.0251\\
0.0727  \!\!\!&\!\!\! -0.0286  \!\!\!&\!\!\!  0.0512  \!\!\!&\!\!\!  0.0377  \!\!\!&\!\!\!  0.0499  \!\!\!&\!\!\!  0.0251  \!\!\!&\!\!\! -0.0462 \!\!\!&\!\!\!  -0.0207  \!\!\!&\!\!\!  0.0017  \!\!\!&\!\!\!  0.0097 \!\!\!&\!\!\!   0.0916  \!\!\!&\!\!\! -0.0750\\
0.0024   \!\!\!&\!\!\! 0.0028  \!\!\!&\!\!\! -0.0012 \!\!\!&\!\!\!   0.0109   \!\!\!&\!\!\! 0.0025  \!\!\!&\!\!\!  0.0001 \!\!\!&\!\!\!  -0.0012 \!\!\!&\!\!\!  -0.0033 \!\!\!&\!\!\!  -0.0027  \!\!\!&\!\!\!  0.0043  \!\!\!&\!\!\!  0.0104  \!\!\!&\!\!\! -0.0033\\
-0.0408  \!\!\!&\!\!\!  0.0169 \!\!\!&\!\!\!  -0.0262  \!\!\!&\!\!\! -0.0225  \!\!\!&\!\!\! -0.0259  \!\!\!&\!\!\! -0.0135 \!\!\!&\!\!\!   0.0280 \!\!\!&\!\!\!   0.0146  \!\!\!&\!\!\! -0.0030  \!\!\!&\!\!\! -0.0078 \!\!\!&\!\!\!  -0.0490 \!\!\!&\!\!\!   0.0473\\
0.0079 \!\!\!&\!\!\!  -0.0126  \!\!\!&\!\!\!  0.0067 \!\!\!&\!\!\!  -0.0108  \!\!\!&\!\!\!  0.0020  \!\!\!&\!\!\!  0.0026 \!\!\!&\!\!\!  -0.0113 \!\!\!&\!\!\!  -0.0022 \!\!\!&\!\!\!   0.0071  \!\!\!&\!\!\! -0.0007 \!\!\!&\!\!\!  -0.0015 \!\!\!&\!\!\!  -0.0186\\
0.0186  \!\!\!&\!\!\! -0.0063  \!\!\!&\!\!\!  0.0097 \!\!\!&\!\!\!   0.0168  \!\!\!&\!\!\!  0.0130  \!\!\!&\!\!\!  0.0053  \!\!\!&\!\!\! -0.0146  \!\!\!&\!\!\! -0.0087 \!\!\!&\!\!\!  -0.0011  \!\!\!&\!\!\!  0.0076  \!\!\!&\!\!\!  0.0301 \!\!\!&\!\!\!  -0.0256\\
0.1159 \!\!\!&\!\!\!  -0.0427 \!\!\!&\!\!\!   0.0801 \!\!\!&\!\!\!   0.0683  \!\!\!&\!\!\!  0.0817 \!\!\!&\!\!\!   0.0396  \!\!\!&\!\!\! -0.0732 \!\!\!&\!\!\!  -0.0340 \!\!\!&\!\!\!  -0.0004  \!\!\!&\!\!\!  0.0179 \!\!\!&\!\!\!   0.1552 \!\!\!&\!\!\!  -0.1189
\end{array}} \!\!\! \right].\label{dama2}
\end{align}
\end{figure*}

We take the weighted adjacency matrix $\textbf{B} \triangleq [w_{i,j}]$ as
\begin{align}
\textbf{B} = \left[ {\begin{array}{*{21}{c}}
0    \!\!\!&\!\!\!     0    \!\!\!&\!\!\!     0     \!\!\!&\!\!\!    0     \!\!\!&\!\!\!    0    \!\!\!&\!\!\!     0         \!\!\!&\!\!\! 0    \!\!\!&\!\!\!     0     \!\!\!&\!\!\!    0   \!\!\!&\!\!\!      0   \!\!\!&\!\!\!      0  \!\!\!&\!\!\!  0.4\\
0.5   \!\!\!&\!\!\!      0      \!\!\!&\!\!\!   0    \!\!\!&\!\!\!     0     \!\!\!&\!\!\!    0    \!\!\!&\!\!\!     0   \!\!\!&\!\!\!      0    \!\!\!&\!\!\!     0    \!\!\!&\!\!\!     0   \!\!\!&\!\!\!      0    \!\!\!&\!\!\!     0  \!\!\!&\!\!\!       0\\
0  \!\!\!&\!\!\!  0.6    \!\!\!&\!\!\!     0    \!\!\!&\!\!\!     0    \!\!\!&\!\!\!     0   \!\!\!&\!\!\!      0  \!\!\!&\!\!\!       0  \!\!\!&\!\!\!       0    \!\!\!&\!\!\!     0    \!\!\!&\!\!\!     0     \!\!\!&\!\!\!    0         \!\!\!&\!\!\! 0\\
0     \!\!\!&\!\!\!    0  \!\!\!&\!\!\!  0.7  \!\!\!&\!\!\!       0    \!\!\!&\!\!\!     0     \!\!\!&\!\!\!    0   \!\!\!&\!\!\!      0    \!\!\!&\!\!\!     0   \!\!\!&\!\!\!      0 \!\!\!&\!\!\!        0     \!\!\!&\!\!\!    0         \!\!\!&\!\!\! 0\\
0    \!\!\!&\!\!\!     0   \!\!\!&\!\!\!      0   \!\!\!&\!\!\! 0.1    \!\!\!&\!\!\!     0     \!\!\!&\!\!\!    0    \!\!\!&\!\!\! 0.2    \!\!\!&\!\!\!     0    \!\!\!&\!\!\!     0  \!\!\!&\!\!\!  0.3    \!\!\!&\!\!\!     0    \!\!\!&\!\!\!     0\\
0   \!\!\!&\!\!\!      0    \!\!\!&\!\!\!     0   \!\!\!&\!\!\!      0   \!\!\!&\!\!\! 0.2     \!\!\!&\!\!\!    0    \!\!\!&\!\!\! 0.3   \!\!\!&\!\!\!      0   \!\!\!&\!\!\!      0   \!\!\!&\!\!\!      0     \!\!\!&\!\!\!    0   \!\!\!&\!\!\!      0\\
0   \!\!\!&\!\!\!      0    \!\!\!&\!\!\!     0    \!\!\!&\!\!\!     0     \!\!\!&\!\!\!    0  \!\!\!&\!\!\!  0.5  \!\!\!&\!\!\!       0    \!\!\!&\!\!\!     0     \!\!\!&\!\!\!    0  \!\!\!&\!\!\!       0  \!\!\!&\!\!\!  0.2   \!\!\!&\!\!\!      0\\
0   \!\!\!&\!\!\!      0     \!\!\!&\!\!\!    0  \!\!\!&\!\!\!       0     \!\!\!&\!\!\!    0    \!\!\!&\!\!\!     0  \!\!\!&\!\!\!  0.1      \!\!\!&\!\!\!   0     \!\!\!&\!\!\!    0  \!\!\!&\!\!\!  0.7     \!\!\!&\!\!\!    0   \!\!\!&\!\!\!      0\\
0   \!\!\!&\!\!\!      0    \!\!\!&\!\!\!     0     \!\!\!&\!\!\!    0     \!\!\!&\!\!\!    0   \!\!\!&\!\!\!      0  \!\!\!&\!\!\!       0   \!\!\!&\!\!\! 0.8     \!\!\!&\!\!\!    0      \!\!\!&\!\!\!   0      \!\!\!&\!\!\!   0     \!\!\!&\!\!\!    0\\
0    \!\!\!&\!\!\!     0   \!\!\!&\!\!\!      0      \!\!\!&\!\!\!   0     \!\!\!&\!\!\!    0    \!\!\!&\!\!\!     0  \!\!\!&\!\!\!       0    \!\!\!&\!\!\!     0  \!\!\!&\!\!\!  0.6    \!\!\!&\!\!\!     0     \!\!\!&\!\!\!    0     \!\!\!&\!\!\!    0\\
0    \!\!\!&\!\!\!     0      \!\!\!&\!\!\!   0     \!\!\!&\!\!\!    0     \!\!\!&\!\!\!    0    \!\!\!&\!\!\!     0  \!\!\!&\!\!\!       0     \!\!\!&\!\!\!    0    \!\!\!&\!\!\!     0  \!\!\!&\!\!\!  0.9    \!\!\!&\!\!\!     0    \!\!\!&\!\!\!     0\\
0   \!\!\!&\!\!\!      0     \!\!\!&\!\!\!    0    \!\!\!&\!\!\!     0   \!\!\!&\!\!\!      0  \!\!\!&\!\!\!  0.2   \!\!\!&\!\!\!      0    \!\!\!&\!\!\!     0    \!\!\!&\!\!\!     0   \!\!\!&\!\!\!      0 \!\!\!&\!\!\!   0.5    \!\!\!&\!\!\!     0
\end{array}} \right].\nonumber
\end{align}

We observe and collect the data of involving opinions until the data matrix $\textbf{P}$ is full-rank. The two data matrices constructed via \eqref{defmatb} and \eqref{defly} are obtained as \eqref{dama1} and \eqref{dama2}, respectively, by which we obtain $\widetilde{\textbf{W}}$ from \eqref{thm21}, i.e., $\widetilde{\textbf{W}} = \textbf{Q}\textbf{P}^{-1}$,
\begin{align}
\widetilde{\textbf{W}} \!\!=\!\!\!\left[\!\!\!\! {\begin{array}{*{21}{c}}
0.2254 \!\!\!&\!\!\!   0  \!\!\!&\!\!\! 0  \!\!\!&\!\!\!  0  \!\!\!&\!\!\!  0     \!\!\!&\!\!\!    0  \!\!\!&\!\!\! 0    \!\!\!&\!\!\! 0 \!\!\!&\!\!\!  0  \!\!\!&\!\!\! 0  \!\!\!&\!\!\! 0  \!\!\!&\!\!\!  0.4\\
0.5  \!\!\!&\!\!\!  0.051  \!\!\!&\!\!\! 0  \!\!\!&\!\!\! 0  \!\!\!&\!\!\! 0  \!\!\!&\!\!\! 0 \!\!\!&\!\!\!  0  \!\!\!&\!\!\!       0  \!\!\!&\!\!\! 0   \!\!\!&\!\!\!      0  \!\!\!&\!\!\! 0  \!\!\!&\!\!\!  0\\
0 \!\!\!&\!\!\!  0.6 \!\!\!&\!\!\! 0.0506   \!\!\!&\!\!\!      0  \!\!\!&\!\!\!  0 \!\!\!&\!\!\!  0  \!\!\!&\!\!\!       0  \!\!\!&\!\!\!  0 \!\!\!&\!\!\!  0   \!\!\!&\!\!\!      0 \!\!\!&\!\!\!  0   \!\!\!&\!\!\!      0\\
0 \!\!\!&\!\!\!  0  \!\!\!&\!\!\!  0.7  \!\!\!&\!\!\!  0.0699   \!\!\!&\!\!\!      0    \!\!\!&\!\!\!     0 \!\!\!&\!\!\!  0    \!\!\!&\!\!\!     0   \!\!\!&\!\!\!      0   \!\!\!&\!\!\!      0 \!\!\!&\!\!\!  0 \!\!\!&\!\!\!  0\\
0  \!\!\!&\!\!\! 0   \!\!\!&\!\!\! 0  \!\!\!&\!\!\!  0.1 \!\!\!&\!\!\!  0  \!\!\!&\!\!\!  0  \!\!\!&\!\!\!  0.2 \!\!\!&\!\!\!  0 \!\!\!&\!\!\!   0  \!\!\!&\!\!\!  0.3  \!\!\!&\!\!\!  0 \!\!\!&\!\!\!  0\\
0 \!\!\!&\!\!\!  0  \!\!\!&\!\!\!  0 \!\!\!&\!\!\! 0  \!\!\!&\!\!\!  0.2 \!\!\!&\!\!\!   0  \!\!\!&\!\!\!  0.3 \!\!\!&\!\!\!  0  \!\!\!&\!\!\!  0  \!\!\!&\!\!\!  0  \!\!\!&\!\!\!  0 \!\!\!&\!\!\!  0\\
0   \!\!\!&\!\!\!      0  \!\!\!&\!\!\!  0    \!\!\!&\!\!\!    0 \!\!\!&\!\!\!  0  \!\!\!&\!\!\!  0.5   \!\!\!&\!\!\!      0  \!\!\!&\!\!\! 0 \!\!\!&\!\!\!   0     \!\!\!&\!\!\!    0  \!\!\!&\!\!\!  0.2    \!\!\!&\!\!\!     0\\
0  \!\!\!&\!\!\! 0 \!\!\!&\!\!\!  0 \!\!\!&\!\!\!  0 \!\!\!&\!\!\!  0 \!\!\!&\!\!\!   0   \!\!\!&\!\!\! 0.1 \!\!\!&\!\!\!  0  \!\!\!&\!\!\!  0  \!\!\!&\!\!\!  0.7 \!\!\!&\!\!\!   0 \!\!\!&\!\!\!  0\\
0    \!\!\!&\!\!\!     0 \!\!\!&\!\!\!   0 \!\!\!&\!\!\!  0  \!\!\!&\!\!\! 0   \!\!\!&\!\!\! 0  \!\!\!&\!\!\!  0  \!\!\!&\!\!\!  0.8    \!\!\!&\!\!\!     0   \!\!\!&\!\!\! 0  \!\!\!&\!\!\! 0 \!\!\!&\!\!\!  0\\
0  \!\!\!&\!\!\! 0  \!\!\!&\!\!\!  0  \!\!\!&\!\!\!  0  \!\!\!&\!\!\!  0  \!\!\!&\!\!\! 0  \!\!\!&\!\!\! 0  \!\!\!&\!\!\!  0  \!\!\!&\!\!\!  0.6 \!\!\!&\!\!\!  0    \!\!\!&\!\!\!     0  \!\!\!&\!\!\!  0\\
0   \!\!\!&\!\!\!      0  \!\!\!&\!\!\! 0  \!\!\!&\!\!\! 0  \!\!\!&\!\!\! 0 \!\!\!&\!\!\!   0  \!\!\!&\!\!\!  0 \!\!\!&\!\!\!  0  \!\!\!&\!\!\!  0  \!\!\!&\!\!\!  0.9  \!\!\!&\!\!\!  0   \!\!\!&\!\!\!     0\\
0 \!\!\!&\!\!\!  0 \!\!\!&\!\!\!   0  \!\!\!&\!\!\! 0 \!\!\!&\!\!\! 0   \!\!\!&\!\!\! 0.2 \!\!\!&\!\!\!   0   \!\!\!&\!\!\!      0  \!\!\!&\!\!\!  0  \!\!\!&\!\!\!  0  \!\!\!&\!\!\!  0.5 \!\!\!&\!\!\!  0
\end{array}}\!\!\!\! \right]\!\!\!. \nonumber
\end{align}

Through comparing the off-diagonal entries of $\widetilde{\textbf{W}}$  and $\textbf{B}$, we conclude that $[\widetilde{\textbf{W}}]_{i,j} = [\textbf{B}]_{i,j} = w_{ij}$, $\forall i \neq j$, which indicates that the network topology is exactly inferred. Moreover, since $x_{i}(0) \neq 0$ for $i = 1,2,3,4$, by \eqref{thm32} and ${[\widetilde{\textbf{W}}]_{i,i}}, i =1,2,3,4$, we have ${\gamma _1} = \frac{{0.2254}}{{0.7513}} = 0.3,{\gamma _2} = \frac{{0.0510}}{{0.2551}} = 0.2,{\gamma _3} = \frac{{0.0506}}{{0.5060}} = 0.1,{\gamma _4} = \frac{{0.0699}}{{0.6990}} = 0.1$. Thus, the parameters $\gamma_{i}$ of confirmation bias model \eqref{lmodel} in \eqref{lmodl}--\eqref{lmod4} are exactly inferred. Finally, the remaining parameters are obtained via \eqref{thm33}: $({{\beta _1},{\beta _2},{\beta _3},{\beta _4}} ) = ({0.5,0..4,0.3,0.2})$. These computation results demonstrate the effectiveness of exact inference procedure presented in Theorem \ref{thm3}.

\subsection{Approximate Inference}
\subsubsection{Approximation Errors}
We now study the approximate inference in the well-known Krackhardt's advice network~\cite{krackhardt1987cognitive}. Its communication structure is shown in Figure~\ref{mm} where the gray nodes represent 21 managers and the red node denotes the information source $\emph{\emph{I}}$.

We take $w_{i,j}$s as follows:  if manager $i$ is not the follower of $\emph{\emph{I}}$ and asks for advise from neighbor $j$, $w_{i,j} = \frac{1}{\Gamma_{i}}$, where ${{\Gamma}_i}$ is manager $i$'s in-degree; if manager $i$ is the follower of $\emph{\emph{I}}$ and asks for advise from neighbor $j$, $w_{i,j} = \frac{1}{1.125\Gamma_{i}+0.155}$; otherwise, $w_{i,j} = 0$. We assume that the information source value is $u_1 = 0.5$ and the bias models (unknown to the inference algorithm) of the four followers are:
\begin{align}
{{\hat w}_{3,d}}({{x_3}(k)}) &= 0.13 - 0.13\sin( {\left| {{x_3}( k ) - u} \right|}),\nonumber\\
{{\hat w}_{4,d}}({{x_4}(k)}) &= 0.125 - 0.125\sin( {\left| {{x_4}( k ) - u} \right|}),\nonumber\\
{{\hat w}_{19,d}}({{x_{19}}(k)}) &= 0.14\log( {2 - {x_{19}}( k )}),\nonumber\\
{{\hat w}_{20,d}}({{x_{20}}(k)}) &= 0.125\log( {2 - {x_{20}}( k )}). \nonumber
\end{align}

To quantify the topology inference error, we define
\begin{align}
e_{i,j}  \triangleq {E_U}( {{|\breve{w}}_{i,j} - { w_{i,j}|}}),\label{eq:frn10}
\end{align}
where $\breve{w}_{i,j} = [\textbf{Q}_{m,q}\textbf{P}_{m,q}^{-1}]_{i,j}$.

\begin{figure}[http]
\centering{
\includegraphics[scale=0.40]{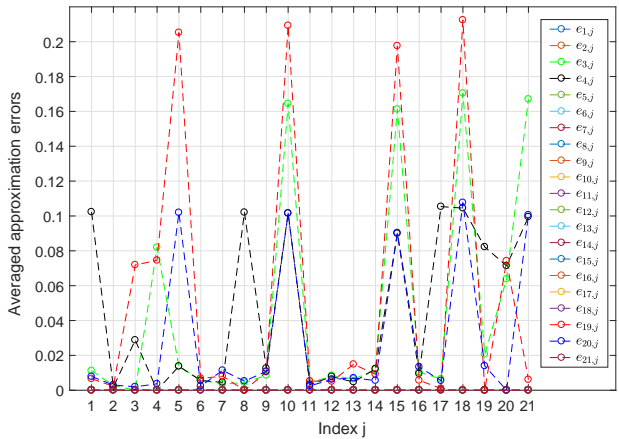}
}
\caption{Average approximation errors of topology.}
\label{bb}
\end{figure}

We  set the observation start time in \eqref{noidefm} and \eqref{lm3rd1} as $m = 2$. We plot $\breve{w}_{i,j}$ averaged  over 1000 runs of innate opinions, randomized uniformly in $[0,1]$ in Figure~\ref{bb}.  We make two observations from Figure \ref{bb}.
\begin{itemize}
  \item Proposed approximate topology inference algorithm infers the  topology from the neighbors to the non-followers of information source, as expected from the theoretical results in Theorem \ref{thm4};
  \item  the approximation error of weighted topology from the neighbors to the non-followers of information source are significantly smaller than those of the followers.
\end{itemize}

The significant approximation errors of followers of information source can be explained by two facts:
\begin{itemize}
  \item The confirmation bias model is unknown to the inference algorithm.
  \item The information source has a direct influence in the resistance parameters of her followers, which then influence the approximate errors.
\end{itemize}

\subsubsection{Length of Recorded Data of Evolving Opinions}

\begin{figure}[!t]
\centering
\subfigure{\includegraphics[scale=0.40]{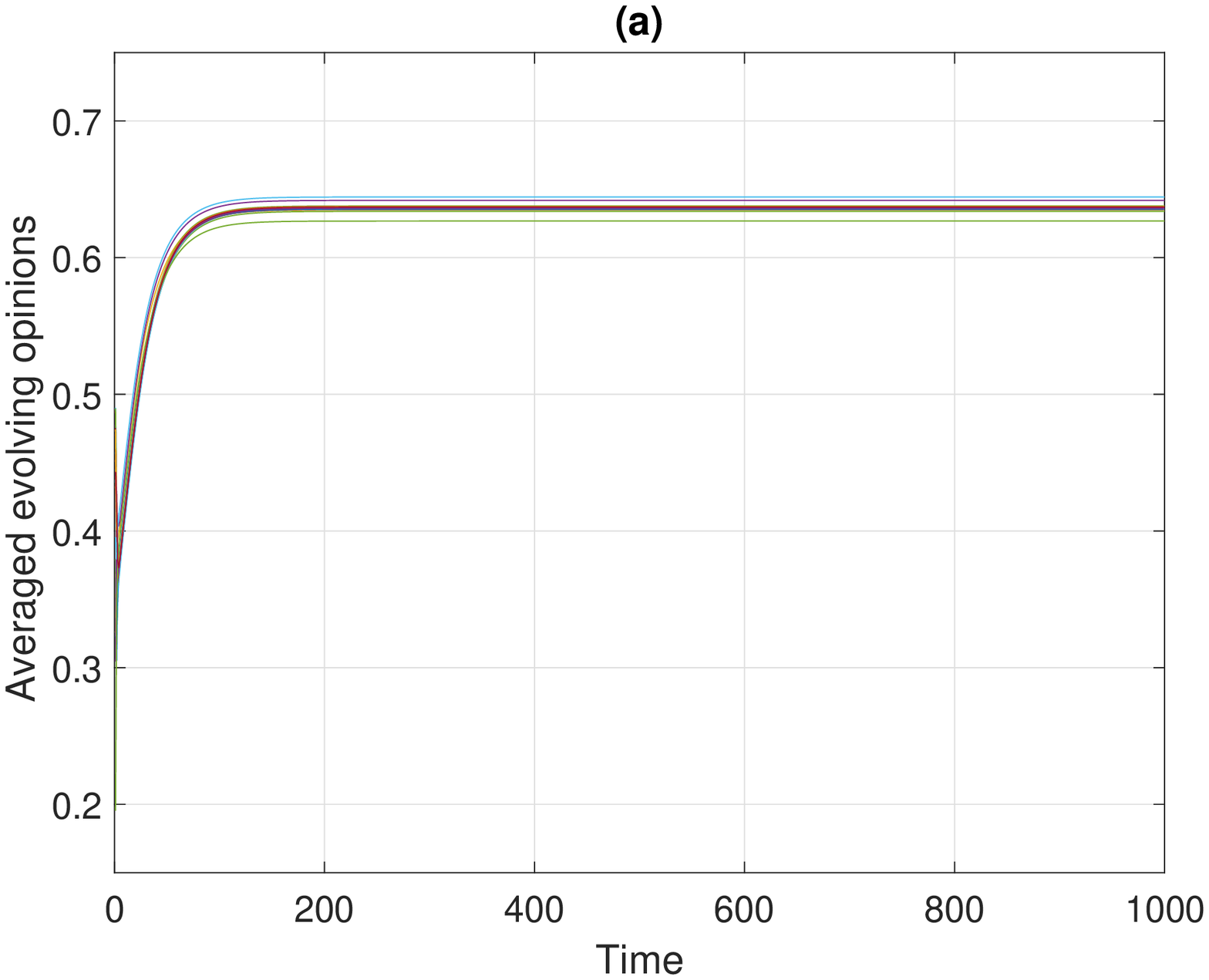}}\\
\subfigure{\includegraphics[scale=0.40]{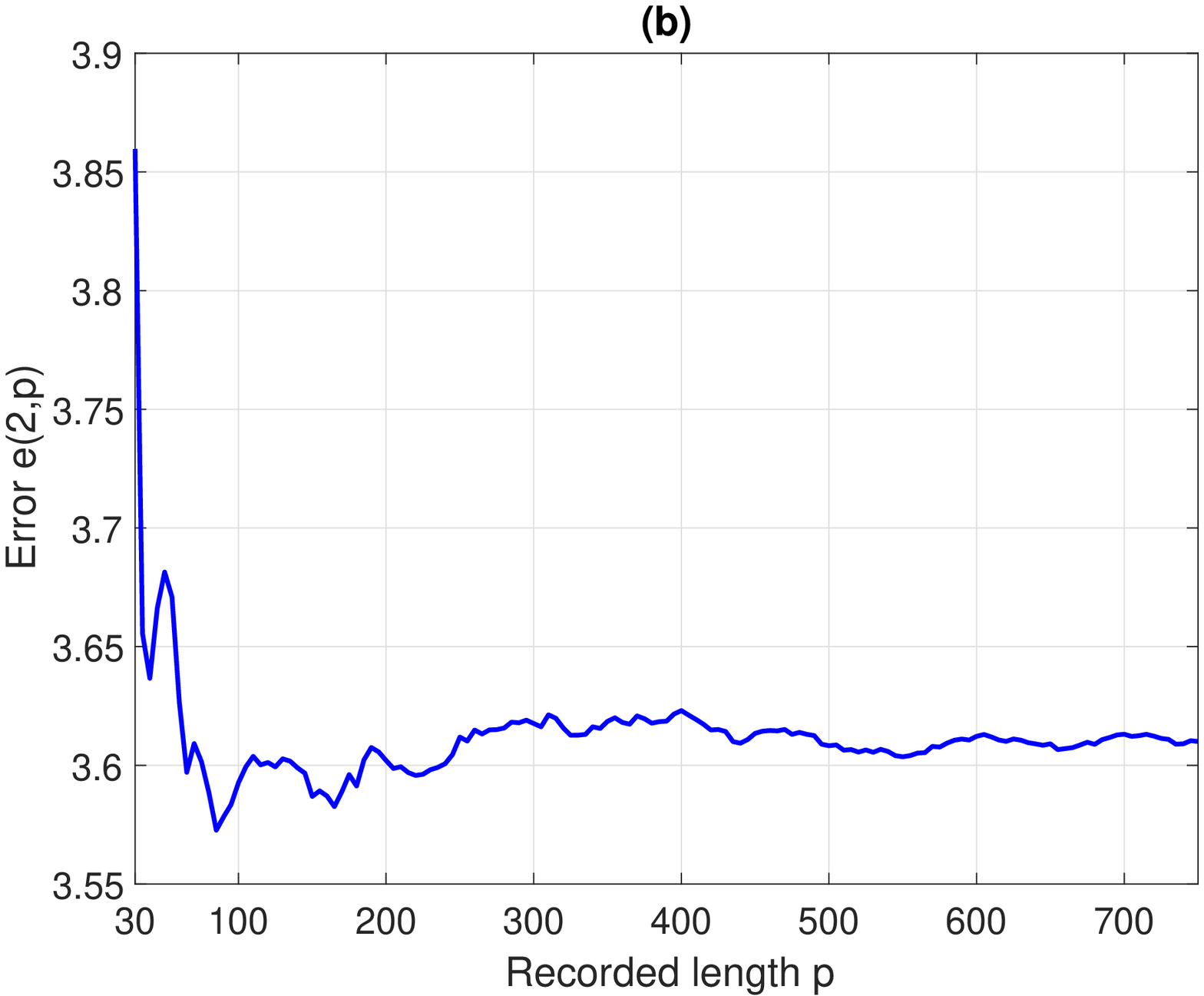}}
\caption{(a) Trajectories of average evolving opinions, (b) approximation error $e({m,p})$ versus $p$.}
\label{fig:tpi1}
\end{figure}

To analyze the influence of length of recorded data, i.e., $p-m$, on the approximation errors, we define the metric:
\begin{align}
e({m,p}) \triangleq \sum\limits_{i \ne j \in \mathbb{V}} {{{{[{E_U}({{|\breve{\textbf{W}}_{m,p}} - \breve{\textbf{W}}}|)]}_{i,j}}}}. \label{mec}
\end{align}
where $\breve{\textbf{W}}_{m,p} = \breve{\textbf{Q}}_{m,p}\breve{\textbf{P}}^{-1}_{m,p}$. With 1000 random samples of innate opinions, we present the average opinion trajectories in Figure~\ref{fig:tpi1} (a).

To guarantee that the matrix \eqref{noidefm} satisfies $\text{rank}(\textbf{P}_{m,p}) = 21$, $m$ and $p$ must satisfy $p - m \geq 21$ and $m \geq 2$. We fix $m = 2$. The approximation error $e({m,p})$ in term of recorded length $p$ are given in Figure~\ref{fig:tpi1} (b), which shows that increasing the length of recoded data of evolving opinion does not necessarily result in more accurate approximation, and as the recorded length increases, the approximate errors measured by \eqref{mec} do not change significantly. This can be explained by the average opinion trajectories in Figure~\ref{fig:tpi1} (a) that after time $k = 500$, the dynamics nearly reaches its steady state. Therefore, $\textbf{x}( {k + 1}) - \textbf{x}( k ) \approx {\mathbf{0}}$ for $k \geq 22$, which with the definition \eqref{noidefm} implies that the recorded data around the equilibrium point has insignificant contribution to $\textbf{P}_{2,p}$ for $p \geq 500$.

\section{Conclusion}
In this paper, we have analyzed the problem of joint inference of network topology and confirmation bias parameters from opinion dynamics. We have analyzed three inference problems: i) piece-wise bias with controlled information sources (stubborn individuals) ii) no bias and no controlled information sources iii) unknown bias model with uncontrolled information sources. We have characterized conditions for exact inference, when possible (for the first two cases) and approximate it when not (for the final case). Numerical simulations performed over a toy network as well as the well-known Krackhardt's advice network  suggest the effectiveness of the obtained theoretical results.

 We consider this work as a part of a comprehensive exploration of several interesting but challenging research problems in this area. Admittedly,  proposed methods in this paper are not designed to work for large-scale networks (e.g., ones in the scale of billion  nodes) since we use matrix computations regularly in our algorithms. Designing computationally efficient inference algorithms with a focus on large-scale networks is left as a part of future work.

Another future research direction is the analysis of the impact of an adversary (or competitor) who is controlling a subset of information sources on the topology inference. This consideration is expected to bring a trade-off between the performances of  the topology inference and opinion evolution control. Some preliminary results in this direction can be found in \cite{asilomar}.

\section{Acknowledgement }
This work is supported by the Interdisciplinary Collaboration Grant 2019, ``Mathematical Models of Media Bias, Polarization and Misinformation Spread Over Networks", Binghamton University-SUNY, and in part by the NSF under the grant CCF \#1910715.

We would like to thank the reviewers for their comments that helped to clarify our results. Particularly, Remark 6 and comments on the computational complexity of the proposed methods are inspired by the reviewer comments.

\appendices
\vspace{-0.00cm}
\section*{Appendix A: Proof of Proposition \ref{prop1}}
The trajectory $\textbf{x}(k)$ is obtained from \eqref{eq:hh1}:
\begin{align}
\textbf{x}(k) = ({{\textbf{W}^k} + \sum\limits_{l = 0}^{k - 1} {{\textbf{W}^l}} \textbf{A}})\textbf{x}(0), \quad k = 1, \ldots, m.\label{traj}
\end{align}

We note that \eqref{eq:hh2} can equivalently be expressed as
\begin{align}
\textbf{x}(k + 1) = \textbf{A}\textbf{x}(0) + \textbf{W}\textbf{x}(k) + \widehat{\textbf{A}}\textbf{x}(0) + \widehat{\textbf{W}}\textbf{x}(k),\label{reeq2}
\end{align}
where $\widehat{\textbf{A}}$ and $\widehat{\textbf{W}}$ are defined in \eqref{defma}. We conclude from \eqref{eq:hh1} and \eqref{reeq2} that $\textbf{x}( k )$ is the solution to both \eqref{eq:hh1} and  \eqref{eq:hh2} if and only if $\widehat{\textbf{A}}\textbf{x}(0) + \widehat{\textbf{W}}\textbf{x}(k) = {\mathbf{0}},\forall k = 0, 1, \ldots, m-1$, which, in conjunction with \eqref{traj}, is equivalent to
\begin{align}
({\widehat{\textbf{A}} + \widehat{\textbf{W}}})\textbf{x}(0) &= {\textbf{0}},\label{iff1}\\
({\widehat{\textbf{A}} + \widehat{\textbf{W}}({{\textbf{W}^k} + \sum\limits_{l = 0}^{k - 1} {{\textbf{W}^l}} \textbf{A}})})\textbf{x}(0) &= {\textbf{0}},k = 1, \ldots, m-1.\label{iff2}
\end{align}

With the consideration of the definitions \eqref{defma} and \eqref{defla}, the condition \eqref{iffc} is equivalently described by \eqref{iff1} and
\begin{align}
&\widehat{\textbf{A}}{\textbf{A}^l}\textbf{L}\textbf{x}(0) \nonumber \\
&= \widehat{\textbf{W}}{\textbf{W}^l}({\textbf{A} + \textbf{W} - \textbf{I}})\textbf{x}(0) = \textbf{0}, l = 0, 1, \ldots, m-1.\label{edes}
\end{align}
Adding \eqref{edes} with $l = 0$ to \eqref{iff1} yields
\begin{align}
({\widehat{\textbf{A}} + \widehat{\textbf{W}}({\textbf{W} + \textbf{A}})})\textbf{x}(0) = \textbf{0}.\label{add1}
\end{align}
Adding \eqref{edes} with $l = k$ to \eqref{add1} results in
\begin{align}
({\widehat{\textbf{A}} + \widehat{\textbf{W}}({{\textbf{W}^{k+1}} + \sum\limits_{l = 0}^{k} {{\textbf{W}^l}} \textbf{A}})})\textbf{x}(0) = {\mathbf{0}},k = 0, \ldots, m-1 \nonumber
\end{align}
which is equivalent to \eqref{iff2}.

\section*{Appendix B: Proof of Theorem \ref{thm1}}
We first note that there  exist  matrices $\widehat{\textbf{A}} \in \mathbb{R}^{n \times n}$ and $\widehat{\textbf{W}} \in \mathbb{R}^{{n \times n}}$, see e.g., $\widehat{\textbf{A}} = \widehat{\textbf{W}} = \mathbf{O}$ (which cannot guarantee the unique inference solution of the parameters of confirmation bias, that is demonstrated in Remark \ref{rem3}), such that \eqref{iffc} holds for any $\textbf{x}(0) \in \mathbb{R}^{n}$.

We let $\textbf{x}(k)$ be the evolving opinion to both \eqref{eq:hh1} and \eqref{eq:hh2} for $k = 1, 2, \ldots, n$. By the relation $\textbf{x}(0) \in {\textbf{L}^{ - 1}}\ker({\widehat{\textbf{O}}})$ implied by Proposition \ref{prop1},  we have $\widehat{\textbf{W}}{\textbf{W}^k}\textbf{L}\textbf{x}(0) = \textbf{0}, k = 0, 1, \ldots, n-1$, which follows from the definition of $\widehat{\textbf{W}}$ in \eqref{defma}:
\begin{align}
{\textbf{W}^{k+1}}\textbf{L}\textbf{x}(0) = \widetilde{\textbf{W}}{\textbf{W}^k}\textbf{L}\textbf{x}(0), k = 0, 1, \ldots, n-1. \label{reiff}
\end{align}

We note that \eqref{reiff} implies ${\textbf{W}^k}\textbf{L}\textbf{x}(0) = \widetilde{\textbf{W}}{\textbf{W}^{k - 1}}\textbf{L}\textbf{x}(0)$, $k = 1, 2, \ldots, n$, substituting which back into \eqref{reiff} yields ${\textbf{W}^{k + 1}}\textbf{L}\textbf{x}( 0 ) = {\widetilde{\textbf{W}}^2}{\textbf{W}^{k - 1}}\textbf{L}\textbf{x}(0)$. Following the same analysis, we have
${\textbf{W}^k}\textbf{L}\textbf{x}( 0 ) = {{\widetilde{\textbf{W}}^k}}\textbf{L}\textbf{x}(0), k = 1, \ldots, n$, by which we obtain
\begin{align}
&\textbf{W}[{\textbf{L}\textbf{x}(0),\textbf{W}\textbf{L}\textbf{x}(0), \ldots ,{\textbf{W}^{n - 1}}\textbf{L}\textbf{x}(0)}] \nonumber\\
& = \widetilde{\textbf{W}}[{\textbf{L}\textbf{x}(0),\widetilde{\textbf{W}}\textbf{L}\textbf{x}(0), \ldots ,{{\widetilde{\textbf{W}}}^{n - 1}}\textbf{L}\textbf{x}(0)}] \nonumber\\
& = \widetilde{\textbf{W}}[{\textbf{L}\textbf{x}(0),\textbf{W}\textbf{L}\textbf{x}(0), \ldots ,{\textbf{W}^{n - 1}}\textbf{L}\textbf{x}(0)}]. \label{reiff1}
\end{align}

Condition \eqref{suff1} implies that  $[{\textbf{L}\textbf{x}(0), \ldots ,{\textbf{W}^{n - 1}}\textbf{L}\textbf{x}(0)}]$ is invertible. Thus, from \eqref{reiff1} we obtain
\begin{align}
\textbf{W} &= \widetilde{\textbf{W}}[{\textbf{L}\textbf{x}(0), \ldots ,{\textbf{W}^{n - 1}}\textbf{L}\textbf{x}(0)}]{[{\textbf{L}\textbf{x}(0), \ldots ,{\textbf{W}^{n - 1}}\textbf{L}\textbf{x}(0)}]^{ - 1}} \nonumber\\
&= \widetilde{\textbf{W}}. \label{reiff2}
\end{align}

Let us denote the weighted adjacency matrix of social network by $[\textbf{M}]_{i,j} \triangleq \begin{cases}
		0, &i = j \in \mathbb{V}\\
		w_{i,j}, &i \neq j \in \mathbb{V}.
	\end{cases}$ Hence, the encoded matrix $\textbf{W}$ in \eqref{eq:dm1} can be rewritten as
\begin{align}
\textbf{W} = \textbf{M} + \text{diag}\left\{ {{x_1}( 0){\gamma _1}, \ldots ,{x_n}(0){\gamma _n}} \right\}.\label{reiff3}
\end{align}
Similarly, $\widetilde{\textbf{W}}$ can equivalently be expressed as
\begin{align}
\widetilde{\textbf{W}} = \widetilde{\textbf{M}} + \text{diag}\left\{ {{x_1}(0){\tilde{\gamma}_1}, \ldots ,{x_n}( 0){\tilde{\gamma }_n}} \right\}.\label{reiff4}
\end{align}
Since $\left[ \textbf{M} \right]_{i,i} = 0$ for $\forall i \in \mathbb{V}$,  \eqref{reiff2}, in conjunction with \eqref{reiff3} and \eqref{reiff4}, implies
\begin{subequations}
\begin{align}
\textbf{M} &= \widetilde{\textbf{M}}, \label{res1}\\
{x_i}(0){\gamma _i} &= {x_i}(0){{\tilde \gamma }_i},\forall i \in \mathbb{V}.\label{res2}
\end{align}\label{res}
\end{subequations}
Under the condition \eqref{suff2}, \eqref{res2} implies ${\gamma _i} = {{\tilde \gamma }_i},\forall i \in \mathbb{V}$. Therefore, we conclude from \eqref{res} that if the conditions \eqref{suff1} and \eqref{suff2} hold, the weighted network topology encoded in the weighted adjacency matrix $\textbf{M}$, and the communication topology from information source $\mathrm{u}$ to individuals, and the bias parameters $\gamma_{i}$ encoded in the matrix $\text{diag}\left\{ {{x_1}( 0 ){\gamma _1}, \ldots ,{x_n}( 0 ){\gamma _n}} \right\}$ are inferred uniquely.

We next infer the bias parameters $\beta_{i}$  encoded in the matrix $\textbf{A}$. We note that the other condition implied by \eqref{iffc} is $({\widehat{\textbf{A}} + \widehat{\textbf{W}}})\textbf{x}(0) = \textbf{0}$. Due to \eqref{defma}, we have $({\textbf{A} + \textbf{W}})\textbf{x}(0) = ({\widetilde{\textbf{A}} + \widetilde{\textbf{W}}})\textbf{x}(0)$, which follows from \eqref{reiff2} and \eqref{eq:dmk} that ${\beta _i}{x_i}(0) = {\tilde \beta _i}{x_i}(0),\forall i \in \mathbb{V}$. Thus, the parameters $\beta_{i}$ can be inferred uniquely via considering \eqref{suff2}.

Finally, we conclude that under the conditions \eqref{suff1} and \eqref{suff2}, $w_{i,j} = \tilde{w}_{i,j}, i \neq j \in \mathbb{V}$, and $\beta_{i} = \tilde{\beta}_{i}$ and $\gamma_{i} = \tilde{\gamma}_{i}, i \in \mathbb{V}$.
Conversely, if there exists any $w_{ij} \neq \tilde{w}_{i,j}, i \neq j \in \mathbb{V}$, or $\beta_{i} \neq \tilde{\beta}_{i}, i \in \mathbb{V}$, or $\gamma_{i} \neq \tilde{\gamma}_{i}, i \in \mathbb{V}$, such that $\textbf{x}(0) \notin {\textbf{L}^{ - 1}}\ker({\widehat{\textbf{O}}})$. Consequently, $\textbf{x}(0) \notin  {\textbf{L}^{ - 1}}\ker({\widehat{\textbf{O}}}) \cap \ker ({\widehat{\textbf{A}} + \widehat{\textbf{W}}}).$ By Corollary \ref{cor1}, the inference problem is not solvable in this scenario.

\section*{Appendix C: Proof of Lemma \ref{lem1}}
It follows from the trajectory \eqref{traj} that
\begin{align}
 \textbf{x}(k + 1) - \textbf{x}(k) = {\textbf{W}^k}\textbf{L}\textbf{x}(0),k \in {\mathbb{N}_0},\label{defmat2b}
\end{align}
where $\textbf{L}$ is given by \eqref{defla}. Equality \eqref{lyapunovb} is obtained via considering the definition \eqref{defmatb} and the relation \eqref{defmat2b}:
\begin{align}
\textbf{W}\textbf{P} &= \sum\limits_{k = 0}^{m - 1} {{\textbf{W}^{k + 1}}\textbf{L}\textbf{x}(0){{( {\textbf{x}(k + 1) - \textbf{x}(k)})}^\top}} \nonumber\\
& = \sum\limits_{k = 0}^{m - 1} {({\textbf{x}(k + 2) - \textbf{x}(k + 1)}){{({\textbf{x}(k + 1) - \textbf{x}(k)})^\top}}}. \nonumber
\end{align}

\section*{Appendix D: Proof of Lemma \ref{lem2}}
We let $\textbf{w} \in \ker(\textbf{P})$. From \eqref{defmatb}, we have
\begin{align}
{\textbf{w}^\top}\textbf{P}\textbf{w} \!=\! \sum\limits_{k = 0}^{m - 1} {{\textbf{w}^\top}(\textbf{x}(k \!+\! 1) \!-\! \textbf{x}(k)){{(\textbf{x}(k \!+\! 1) \!-\! \textbf{x}(k))}^\top}} \textbf{w} \!=\! 0, \nonumber
\end{align}
by which we obtain
\begin{align}
(\textbf{x}(k + 1) - \textbf{x}(k))^\top{\textbf{w}} &= 0,~~\forall k = 0,m - 1. \label{plem2}
\end{align}

We conclude from the trajectory \eqref{defmat2b} that \eqref{plem2} equals
\begin{align}
\textbf{x}^\top(0)({\textbf{W}^k}\textbf{L})^\top{\textbf{w}} = 0,\forall k = 0,m - 1. \label{plem4}
\end{align}We note that \eqref{plem4} is equivalent to
\begin{align}
\textbf{w} \in \ker ([{\textbf{L}\textbf{x}(0),\textbf{W}\textbf{L}\textbf{x}(0), \ldots ,{\textbf{W}^{m - 1}}\textbf{L}\textbf{x}(0)} ]^\top), \nonumber
\end{align}
by which, we obtain \eqref{reslem2}.

\section*{Appendix E: Proof of Corollary \ref{cor2}}
For $i \in \mathbb{V}$, with the consideration of \eqref{eq:dmk} and \eqref{eq:dm1}, \eqref{eq:hh1} can be written as
\begin{align}
{x_i}({k + 1}) =({1 - \sum\limits_{j \ne i \in \mathbb{V}} {{{[\textbf{W}]}_{i,j}}}  - {\beta _i}}){x_i}(0) + \sum\limits_{j \in \mathbb{V}} {{{[\textbf{W}]}_{i,j}}} {x_j}(k), \nonumber
\end{align}
which directly yields \eqref{confirp}.

\section*{Appendix F: Proof of Theorem \ref{thm2}}
We first prove the necessary condition. We assume that there exists a matrix $\textbf{W}  \neq \widetilde{\textbf{W}}  \in \mathbb{R}^{n \times n}$ such that
\begin{align}
\textbf{W}\textbf{P} = \textbf{Q}. \label{pthm21}
\end{align}Combining \eqref{thm21} and \eqref{pthm21} we arrive at
\begin{align}
(\textbf{W} - \widetilde{\textbf{W}} )\textbf{P} = \textbf{O}. \label{pthm22}
\end{align}

Since the matrix \eqref{defmatb} is symmetric, there exists an orthogonal matrix $\textbf{V}$ such that $\textbf{P} = \textbf{V}\Delta {\textbf{V}^\top}$, where
\begin{align}
\Delta  = \left[ {\begin{array}{*{20}{c}}
\Lambda &{{\textbf{O}}}\\
{\textbf{O}}&{\textbf{O}}
\end{array}} \right]. \label{pthm23}
\end{align}
with $\Lambda$ being a full-rank diagonal matrix. Then, \eqref{pthm22} is written as $(\textbf{W} - \widetilde{\textbf{W}} )\textbf{V}\Delta \textbf{V}^\top  = \textbf{O}$, pre-and post-multiplying which by $\textbf{V}^\top$ and $\textbf{V}$, respectively, yields
\begin{align}
{{\textbf{V}^\top}(\textbf{W} - \widetilde{\textbf{W}} )\textbf{V}}\Delta  = \textbf{O}. \label{pthm24}
\end{align}

Without loss of generality, we let ${\textbf{V}^\top}(\textbf{W} - \widetilde{\textbf{W}})\textbf{V} = \left[ {\begin{array}{*{20}{c}}
{{\textbf{X}_{11}}}&{{\textbf{X}_{12}}}\\
{{\textbf{X}_{21}}}&{{\textbf{X}_{22}}}
\end{array}} \right]$, substituting which into \eqref{pthm24} with the consideration of \eqref{pthm23} results in
\begin{align}
\left[ {\begin{array}{*{20}{c}}
{\textbf{X}_{11}\Lambda}&{\textbf{O}}\\
{{\textbf{X}_{21}}\Lambda }&{\textbf{O}}
\end{array}} \right] = \textbf{O}, \nonumber
\end{align}
which implies ${\textbf{X}_{11}} = \textbf{O}$ and ${\textbf{X}_{21}} = \textbf{O}$, since $\Lambda$ is a diagonal matrix. Thus,
\begin{align}
{\textbf{V}^\top}(\textbf{W} - \widetilde{\textbf{W}})\textbf{V} = \left[ {\begin{array}{*{20}{c}}
{\textbf{O}}&{{\textbf{X}_{12}}}\\
{\textbf{O}}&{{\textbf{X}_{22}}}
\end{array}} \right]. \nonumber
\end{align}
Then, it can be verified from \eqref{pthm23} that $\Delta ({{\textbf{V}^\top}{{(\textbf{W} - \widetilde{\textbf{W}})^\top}}\textbf{V}}) = \textbf{O}$, pre-and post-multiplying which by $\textbf{V}$ and $\textbf{V}^\top$, respectively, yields $\textbf{P}{(\textbf{W} - \widetilde{\textbf{W}})^\top} = \textbf{O}$, which implies that the columns of matrix ${(\textbf{W} - \widetilde{\textbf{W}})^\top}$ belong to $\ker(\textbf{P})$. By Lemma~\ref{lem2}, we have
\begin{align}
[{\textbf{L}\textbf{x}(0),\textbf{W}\textbf{L}\textbf{x}(0), \ldots ,{\textbf{W}^{m - 1}}\textbf{L}\textbf{x}(0)}]^{\top}(\textbf{W} - \widetilde{\textbf{W}})^\top = \textbf{O}, \nonumber
\end{align}
which is equivalent to ${\textbf{x}^\top}(0){( {{\textbf{W}^k}\textbf{L}})^\top}{(\textbf{W} - \widetilde{\textbf{W}})^ \top } = \textbf{0}^\top, k = 0, 1, \ldots, m-1$, which further implies that
\begin{align}
(\textbf{W} - \widetilde{\textbf{W}}){\textbf{W}^k}\textbf{L}\textbf{x}(0) = \textbf{0}, k = 0,1, \ldots ,m - 1.\label{pthm25}
\end{align}

We note that \eqref{pthm25} can be equivalently expressed as $\textbf{x}(0) \in {\textbf{L}^{ - 1}}\ker({\widehat{\textbf{O}}_{m}})$. Moreover, we can set $\widehat{\textbf{A}} = -\widehat{\textbf{W}}$, such that $\textbf{x}(0) \in \ker ({\widehat{\textbf{A}} + \widehat{\textbf{W}}} )$  holds. Here, we conclude $\textbf{x}(0) \in  {\textbf{L}^{ - 1}}\ker({\widehat{\textbf{O}}}) \cap \ker ({\widehat{\textbf{A}} + \widehat{\textbf{W}}} )$, which follows from Corollary~\ref{cor1} that the inference problem is not feasible.

To prove the sufficient condition, we assume the inference problem is not feasible. By Corollary~\ref{cor1}, we have $\textbf{x}(0)$ $\in$  ${\textbf{L}^{ - 1}}\ker({\widehat{\textbf{O}}})$ with $\widehat{\textbf{O}}$ given by \eqref{defmacc}, which, in conjunction with the \eqref{defma},  results in $({\widetilde{\textbf{W}} - \textbf{W}})[{\textbf{L}\textbf{x}(0 ), \textbf{W}\textbf{L}\textbf{x}( 0 ), \ldots ,{\textbf{W}^{m - 1}}\textbf{L}\textbf{x}(0)} ] = \textbf{O}$. Then, by Lemma~\ref{lem2} we have $({\widetilde{\textbf{W}} - \textbf{W}})\textbf{P} = \textbf{O}$. Consequently, $\widetilde{\textbf{W}} \textbf{P} = \textbf{W} \textbf{P}$. Thus, the matrix $\widetilde{\textbf{W}}$ in \eqref{thm21} is not unique.

\section*{Appendix G: Proof of Theorem \ref{thm3}}
By Lemma~\ref{lem2}, $\text{rank}(\textbf{P}) = n$ implies \eqref{suff1}. Since \eqref{suff2} is assumed to hold as well, by Theorem~\ref{thm1} the inference problem is solvable. Moreover, by Theorem~\ref{thm2} there exists a unique $\widetilde{\textbf{W}}$, which follows from \eqref{thm21} as: $\widetilde{\textbf{W}} = \textbf{Q}\textbf{P}^{-1} = \textbf{W}$. Furthermore, the computation of network topology \eqref{thm31}, and the bias parameters $\gamma_{i}$  \eqref{thm32}  follows from \eqref{eq:dm1}. The remaining parameters $\beta_{i}$ follows from Corollary~\ref{cor2} with the consideration of \eqref{suff2}, which determines its uniqueness.

\section*{Appendix H: Proof of Lemma \ref{lem3}}
From \eqref{nsd}, we obtain
\begin{align}
{\textbf{x}}(k + 1) = ({{\breve{\textbf{W}}^{k + 1}} + \sum\limits_{z = 0}^k {{\breve{\textbf{W}}^z}\breve{\textbf{A}}(k - z)} }){\textbf{x}}(0),k \in {\mathbb{N}_0} \nonumber
\end{align}
by which we have
\begin{align}
&{\textbf{x}}(k + 1) - {\textbf{x}}(k) \nonumber\\
&= ({\breve{\textbf{A}}(k) - \breve{\textbf{A}}({k - 1})}){\textbf{x}}(0) + \breve{\textbf{W}}({\textbf{x}(k) - {\textbf{x}}(k - 1)}),\label{plm31}\\
&{\textbf{x}}(k + 2) - {\textbf{x}}(k + 1) \nonumber \\
    &= ({\breve{\textbf{A}}({k + 1}) - \breve{\textbf{A}}(k)}){\textbf{x}}(0) + \breve{\textbf{W}}({\breve{\textbf{A}}(k) - \breve{\textbf{A}}({k - 1})}){\textbf{x}}(0) \nonumber\\
&\hspace{3.0cm} + {\breve{\textbf{W}}^2}({{\textbf{x}}(k) - {\textbf{x}}(k - 1)}), k \in \mathbb{N}. \label{plm32}
\end{align}

Pre-multiplying \eqref{plm31} by $\breve{\textbf{W}}$ and considering \eqref{plm32} yields
\begin{align}
&\breve{\textbf{W}}({{\textbf{x}}(k + 1) - {\textbf{x}}(k)}) \nonumber\\
& = \breve{\textbf{W}}({\breve{\textbf{A}}(k) - \breve{\textbf{A}}({k - 1})}){\textbf{x}}(0) + {\breve{\textbf{W}}^2}({{\textbf{x}}(k) - {\textbf{x}}(k - 1)}) \nonumber\\
& = {\textbf{x}}(k + 2) - {\textbf{x}}(k + 1) - ({\breve{\textbf{A}}({k + 1}) - \breve{\textbf{A}}(k))}){\textbf{x}}(0). \label{plm33}
\end{align}
It can be verified  from \eqref{noidefm} that  \eqref{lm3r} follows from \eqref{plm33}.

\section*{Appendix I: Proof of Theorem \ref{thm4}}
Subtracting $\textbf{W}\breve{\textbf{P}}_{m,p} = \breve{\textbf{Q}}_{m,p}$ from \eqref{lm3r}   yields
\begin{align}
(\breve{\textbf{W}} - \textbf{W})\breve{\textbf{P}}_{m,p} = \breve{\textbf{R}}_{m,p}. \label{diff}
\end{align}

If an individual $\mathrm{v}_{i}$ is not an follower of information sources, i.e, ${{\breve{w}}_{i,d}}({{{x}_i}(k)}) = 0$ for $\forall d \in \mathbb{K}$, from matrix \eqref{eq:ks1} we have
\begin{align}
[\breve{\textbf{A}}(k)]_{i,:} = \text{diag} \{1 - \sum\limits_{j \in \mathbb{V}} {{\breve{w}_{1,j}}}, \ldots, 1 - \sum\limits_{j \in \mathbb{V}} {{\breve{w}_{n,j}}}\},\nonumber
\end{align}
by which, $[ {\breve{\textbf{A}}(k + 1) - \breve{\textbf{A}}(k)}]_{{i,:}} = \textbf{0}^\top$ for $\forall t \in \mathbb{N}_{0}$, which, in conjunction with \eqref{lm3rd2}, yields
\begin{align}
{[\textbf{R}_{m,p}]_{{i,:}}} = \sum\limits_{t = m}^{p} {\textbf{0}^\top {\textbf{x}}(0){{({\textbf{x}}(k + 1) - {\textbf{x}}(k))}^\top }}  = \textbf{0}^\top. \label{pr1}
\end{align}

We note that \eqref{lm3r} is a necessary condition of the uniqueness of inference solution of network topology and confirmation bias. With $\text{rank}(\breve{\textbf{P}}_{m,p}) = n$, we obtain from \eqref{diff} with \eqref{pr1} that ${[ {\breve{\textbf{W}} - \textbf{W}}]_{{i,:}}} = {[\breve{\textbf{R}}_{m,p}]_{{i,:}}}{\textbf{P}^{-1}_{m,p}} = \textbf{0}^\top$, thus,
\begin{align}
{{[\breve{\textbf{W}}]_{i,:}} = {[\textbf{W}]_{{i,:}}}}. \label{pr2}
\end{align}
$\textbf{W}\breve{\textbf{P}}_{m,p}$ $=$ $\breve{\textbf{Q}}_{m,p}$ implies $\textbf{W}$ $=$ $\breve{\textbf{Q}}_{m,p}\breve{\textbf{P}}^{-1}_{m,p}$, thus, \eqref{paifer} is obtained.

If the individual $\mathrm{v}_{i}$ is a follower of an information source but holds no confirmation bias, from \eqref{eq:ks2}  and \eqref{eq:dm1} we have ${[\breve{\textbf{W}}]_{i,i}} \ne {[\textbf{W}]_{i,i}}$, which implies that \eqref{pr2} does not hold.

\bibliographystyle{IEEEtran}
\bibliography{ref}
\end{document}